\documentclass[aps,prx,reprint,superscriptaddress,longbibliography]{revtex4-2}
\usepackage{mathrsfs}
\usepackage{epsfig}
\usepackage{graphicx}
\usepackage{amsfonts}
\usepackage[figuresright]{rotating}
\usepackage{amssymb}
\usepackage{amsmath}
\usepackage{dcolumn}
\usepackage{bm}
\usepackage{xcolor}
\usepackage[colorlinks, citecolor=blue]{hyperref}
\usepackage{amsmath,amssymb,amsfonts,bm}
\usepackage{amsthm,amsmath,amssymb}
\newcommand{\RNum}[1]{\uppercase\expandafter{\romannumeral #1\relax}}
\hypersetup{linkcolor=magenta,urlcolor=blue,citecolor=blue,pdfstartview={FitH},urlcolor=blue}

\newcommand{\tr}{\operatorname{tr}}
\makeatletter
\newsavebox{\@brx}
\newcommand{\llangle}[1][]{\savebox{\@brx}{\(\m@th{#1\langle}\)}%
  \mathopen{\copy\@brx\kern-0.5\wd\@brx\usebox{\@brx}}}
\newcommand{\rrangle}[1][]{\savebox{\@brx}{\(\m@th{#1\rangle}\)}%
  \mathclose{\copy\@brx\kern-0.5\wd\@brx\usebox{\@brx}}}
\makeatother
\def\Re{\text{Re}}

\begin{document}

\title{Spontaneous symmetry breaking in open quantum systems: strong, weak, and strong-to-weak}

% Please use the appropriate macro for each type of information

\author{Ding Gu}
%\thanks{}
%\affiliation{School of Physics, Peking University, Beijing 100871, China}
\affiliation{Institute for Advanced Study, Tsinghua University, Beijing 100084, China}

\author{Zijian Wang}
\affiliation{Institute for Advanced Study, Tsinghua University, Beijing 100084, China}

\author{Zhong Wang}
\email{wangzhongemail@tsinghua.edu.cn}
\affiliation{Institute for Advanced Study, Tsinghua University, Beijing 100084, China}
% \affiliation{Collaborative Innovation Center of Quantum Matter, Beijing 100871, China}


\date{\today}

\begin{abstract}
Depending on the coupling to the environment, symmetries of open quantum systems manifest in two distinct forms, the strong and the weak. We study the spontaneous symmetry breaking among phases with strong symmetry, weak symmetry, and no symmetry. Concrete Liouvillian models with strong and weak symmetry are constructed, and different scenarios of symmetry-breaking transitions are investigated from complementary approaches. It is demonstrated that strong symmetry always spontaneously breaks, either completely, or into the corresponding weak symmetry. For strong $U(1)$ symmetry, we show that strong-to-weak symmetry breaking leads to gapless Goldstone modes dictating diffusion of the symmetry charge in translational invariant systems. We conjecture that this relation among strong-to-weak symmetry breaking, gapless modes, and symmetry-charge diffusion is general for continuous symmetries. It can be interpreted as an ingappability condition for Lindbladian with strong $U(1)$ symmetry and weak translation symmetry, according to which the gapless spectrum does not require non-integer filling. We also investigate the scenario where the strong symmetry breaks completely. In the symmetry-broken phase, we identify an effective Keldysh action with two Goldstone modes, describing fluctuations of the order parameter and diffusive hydrodynamics of the symmetry charge, respectively. We show that weak $U(1)$ SSB naturally leads to time crystalline order. Additionally, it is found that in the presence of strong symmetry, the spontaneous breaking of weak symmetry can depend on the symmetry sector. For a particular model studied here, we uncover a transition from a symmetric phase with a ``Bose surface'' to a symmetry-broken phase with long-range order induced by tuning the filling. It is also shown that, in both weak and strong symmetry cases, the long-range order of $U(1)$ symmetry breaking is possible in spatial dimension $d\geq 3$. Our work outlines the typical scenarios of spontaneous symmetry breaking in open quantum systems, puts forward a theoretical framework to characterize them, and highlights their physical consequences.

\end{abstract}
%\lipsum[1]
%\keywords{}

\maketitle

\section{\label{sec: intro} Introduction}

Symmetries play a fundamental role in condensed matter physics, in which a great variety of phases and phase transitions are described by symmetries and their spontaneous breaking.  Symmetries constrain the form of free energy and, together with the renormalization group, determines the critical behaviors near the phase transition point. Particularly, continuous symmetry breaking gives rise to Goldstone modes \cite{PhysRevLett.4.380,PhysRev.127.965}, such as spin waves and phonons, which render the system gapless.

Nevertheless, the discussion of symmetry and symmetry breaking is usually in the context of ground states or thermal ensembles. For open quantum systems, where coupling to the environment can not be neglected and where systems do not necessarily evolve towards thermal equilibrium, symmetries and their spontaneous breaking are much less explored. Interestingly, symmetries manifest in two distinct forms in open quantum systems: weak and strong \cite{Buca_2012,victor2014symmetries,lieu2020symmetry}. Strong symmetries are similar to symmetries discussed in closed systems, with a conserved symmetry charge during time evolution. In contrast, due to interaction with the environment, weak symmetries do not imply the conservation of a physical symmetry charge. Recently, there has been a growing interest in exploring novel phases of matter in open quantum systems \cite{diehl2008,Sieberer_2016,PhysRevLett.110.195301, PhysRevB.89.134310, PhysRevX.5.011017,zou2023channeling,fan2023diagnostics,bao2023mixed,lee2023quantum,lu2023mixed,hauser2024continuous,wang2023topologically,dai2023steady,liu2024dissipative,wang2023intrinsic,sang2023mixed,chen2023separability,chen2024unconventional,rakovszky2023defining,li2024replica,sieberer2023universality,sohal2024noisy,cheng2024towards,sang2024stability,lu2024disentangling}, where both strong and weak symmetries play significant roles in characterizing these phases \cite{lieu2020symmetry,ma2023average,lee2022symmetry,zhang2022strange,mao2023dissipation,ma2023topological,huang2024mixed,chen2023symmetry,kawabata2023lieb,zhou2023reviving,lessa2024mixed,wang2024anomaly,guo2024locally,ma2024symmetry,xue2024tensor,lessa2024strong,sala2024spontaneous,lessa2024symmetry,pollmann2024highly,xu2024average,zhang2024quantum}.

In this work, we focus on continuous symmetries and their spontaneous breaking in the context of Markovian dynamics governed by Lindblad superoperators, specifically emphasizing the $U(1)$ case. Previous studies of $U(1)$ symmetry breaking in open quantum systems are mainly devoted to driven-dissipative Bose-Einstein condensates \cite{diehl2008,Sieberer_2016,PhysRevLett.110.195301, PhysRevB.89.134310, PhysRevX.5.011017}. Here, by representing strong and weak symmetry operations as superoperators, we are able to systematically analyze how spontaneous symmetry breaking takes place in open quantum systems and their consequences.

For weak $U(1)$ symmetric Liouvillians, spontaneous symmetry breaking is accompanied by Liouvillian gap closing in the thermodynamic limit, with conventional long-range order in the steady states. For strong $U(1)$ symmetric Liouvillians, we find that there is  $U(1)\times U(1)$ symmetry in the doubled space, where one of the $U(1)$ symmetry is identified as weak $U(1)$ symmetry. However, due to the structure of operators governing time evolution, the other $U(1)$ symmetry always spontaneously breaks. This strong-to-weak symmetry breaking is universal, occurring in all dimensions and for any parameter. Consequently, if the Liouvillian is also translational invariant, there will be Goldstone modes, which in the physical space describe the diffusion of conserved symmetry charge, and render the spectrum gapless. In contrast to the conventional Lieb-Schultz-Mattis (LSM) theorem \cite{LIEB1961407,PhysRevLett.84.1535} as well as its recent generalization in open quantum systems \cite{kawabata2023lieb}, we argue that strong continuous symmetry and translational invariance always lead to gapless Liouvillian spectrum, at both integer and non-integer fillings. %Thus, we call it a non-LSM-type ingappability condition for open quantum systems.} 
We note that recently, the strong-to-weak symmetry breaking were studied in Refs. \cite{lee2023quantum,lessa2024strong,sala2024spontaneous} with a rather different focus.   Starting from strong-symmetric mixed states, Refs.~\cite{lee2023quantum,lessa2024strong,sala2024spontaneous} define and characterize strong-to-weak symmetry breaking mainly from an information-theoretic viewpoint. In the present work, we focus on the physical consequences of strong-to-weak symmetry breaking, including the Goldstone modes,  Liouvillian spectrum, relaxation and hydrodynamics. We also provide a field-theoretic understanding of the phenomena.

We also demonstrate new consequnces and scenerios of weak $U(1)$ SSB. It is found that weak $U(1)$ SSB naturally leads to time crystalline order. Additionally, for strong $U(1)$ symmetric Liouvillians, conventional weak $U(1)$ symmetry breaking can exhibit more exotic behaviors. In different symmetry sectors the steady states can possess very different physical properties, and long-range order for weak $U(1)$ spontaneous symmetry breaking occurs only at certain fillings. We argue that this filling-induced transition in the model we study can be understood as being analogous to percolation.

%This can be understood as an “enhanced” Lieb-Schultz-Mattis (LSM) theorem in open quantum systems \cite{LIEB1961407,PhysRevLett.84.1535,PhysRevLett.132.070402}, where the interplay between strong continuous symmetry and translational invariance leads to gapless spectrum at any filling. 
\begin{figure}[htb]
    \centering
    \includegraphics[width=\linewidth]{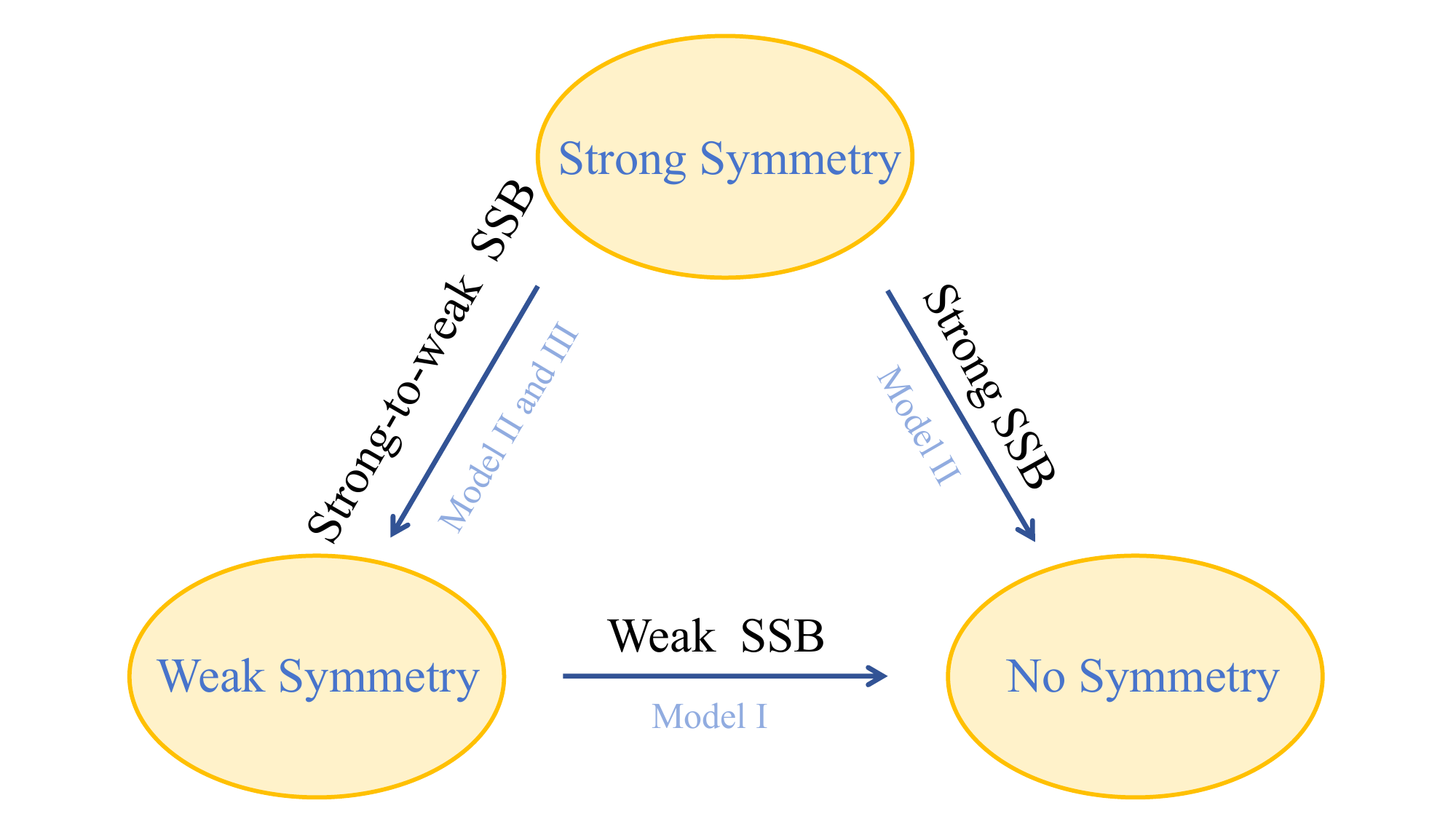}
    \caption{Different spontaneous symmetry breaking processes are discussed in this work, including weak symmetry breaking, strong-to-weak symmetry
breaking, and complete breaking of strong symmetry. Model I with weak $U(1)$ symmetry breaking is studied in Sec.~\ref{sec:weak}. Model II with strong and strong-to-weak $U(1)$ symmetry breaking is studied in Sec.~\ref{sec:strong}. In Sec.~\ref{sec:s-to-w}, we further discuss strong-to-weak symmetry breaking and its connection to the diffusion of the symmetry charge, with model III as an example.}
    \label{fig:models}
\end{figure}

We construct simple spin models with weak and strong $U(1)$ symmetries where different spontaneous symmetry breaking processes are studied, including weak symmetry breaking, strong-to-weak symmetry breaking, and complete breaking of strong symmetry (shown in Fig.~(\ref{fig:models})). For the strong $U(1)$ symmetric model, we find a novel filling-induced transition from a weak $U(1)$ symmetric phase with Bose surface \cite{subir02bs,Varney11bs,Wang2020pattern} to a weak $U(1)$ symmetry broken phase with long-range order. When the strong $U(1)$ symmetry completely breaks, we identify two Goldstone modes in the effective action, describing fluctuations of order parameter and diffusion of the symmetry charge. We also show that long-range order is possible in $d\geq 3$.

The remainder of the article is organized as follows. In Sec.~\ref{sec:sym}, we set up a framework to study weak and strong symmetries, outlining how spontaneous symmetry breaking takes place and the consequences of symmetry breaking.  In Sec.~\ref{sec:weak} and~\ref{sec:strong}, spin models of weak and strong $U(1)$ symmetries are studied, and the symmetry-breaking transitions are illustrated by mean-field, perturbation theory, and field-theoretical analysis. Specifically, for the strong symmetric model, we explain in detail how the change in filling induces the transition from the weak $U(1)$ symmetric phase to the symmetry broken phase. We also derive the effective action describing fluctuations of the order parameter and diffusion of the symmetry charge, where the lower critical dimension is identified. Finally in Sec.~\ref{sec:s-to-w}, we further elaborate on the non-LSM-type ingappability condition, explaining how the interplay between strong continuous symmetry and translational invariance renders the spectrum of Liouvillians gapless in each symmetry sector, and how the gapless modes are connected with diffusion modes of the symmetry charge.

\section{\label{sec:sym} Symmetries in open quantum systems}
In open quantum systems, states are represented by density matrix $\rho$, and for Markovian environments, time evolution is generally governed by Lindblad master equation \cite{10.1063/1.522979,Lindblad}:
\begin{equation}
\dot{\rho} = \mathcal{L}[\rho] = -i[H,\rho]+\sum_{\mu}\gamma_{\mu}(2L_{\mu}\rho L_{\mu}^{\dagger}-\{L_{\mu}^{\dagger}L_{\mu},\rho\}),
\label{eq:master}
\end{equation}
where $H$ is the Hamiltonian that describes unitary evolution, $L_{\mu}$'s are dissipators arising from coupling to the environment, and $\gamma_{\mu}\geq 0$ describes the strength of dissipation associated with each $L_{\mu}$. $\mathcal{L}$ is called the Liouvillian superoperator. As $t\rightarrow\infty$, the system described by Eq.~(\ref{eq:master}) approaches steady state $\rho_{ss}$ with $\mathcal{L}[\rho_{ss}] = 0$, which in general is a mixed state. The Liouvillian spectrum satisfies $0 = \lambda_0\geq\Re\lambda_1\geq\Re\lambda_2\geq\cdots$, where $\lambda_0 = 0$ is the eigenvalue of the steady state $\rho_{ss}$ and the Liovillian gap is defined as $|\Re\lambda_1|$.

There are two types of symmetry in open quantum systems: weak and strong (the names will be explained later). For a symmetry group $G$, with each element $g\in G$ acting as a unitary transformation $U(g)$, the weak symmetry operation is defined as:
\begin{equation}
\mathcal{U}_{\text{w}}(g)[\rho] \equiv U(g)\rho U^{\dagger}(g),
\label{eq:weak}
\end{equation}
and the strong symmetry operation is defined as 
\begin{equation}
\mathcal{U}_{\text{s}}(g)[\rho]\equiv U(g)\rho \quad \text{or}\quad\mathcal{U}_{\text{s}}(g)[\rho]\equiv \rho U^{\dagger}(g). 
\label{eq:stro}
\end{equation}
%A state is weak symmetric if $\mathcal{U}^{w}(g)[\rho] \equiv U(g)\rho U^{\dagger}(g) = 1$, and it is strong symmetric if $\mathcal{U}^s(g)[\rho]\equiv U(g)\rho = e^{i\phi}\rho$.%

A Liouvillian $\mathcal{L}$ respects weak (strong) symmetry $G$ if  $[\mathcal{L}, \mathcal{U}_{\text{w(s)}}(g)] = 0$ for all $g\in G$.  For example, we consider global $U(1)$ symmetry operation $U(\theta) = e^{-iN\theta}$ generated by symmetry charge $N = \sum_i n_i$, with local order parameter $a_i^{\dagger}(a_i)$ satisfying $[n_i,a_i^{\dagger}(a_i)] = a_i^{\dagger}(-a_i)$. If the Hamiltonian $H$ commutes with $N$, and dissipators $L_{\mu}$'s are of the type $a_i$ or $a_i^{\dagger}$, the Liouvillian respects weak $U(1)$ symmetry, but $N$ is not conserved: $\frac{d}{dt}\langle N\rangle \neq 0$ due to the exchange of symmetry charge with the environment. If dissipators are of the type $n_i$ or $a_j^{\dagger}a_i$, the Liouvillian respects strong $U(1)$ symmetry where the total charge $N$ is conserved $\frac{d}{dt}\langle N\rangle = 0$. The names ``strong'' and ``weak'' are related to whether the symmetry charge is conserved or not.

Global symmetries can be spontaneously broken. In the steady states, spontaneous symmetry breaking (SSB) can be detected by long-range order or degeneracy in the thermodynamic limit.  Here we focus on the continuous case $U(1)$. For weak $U(1)$ symmetry, spontaneous symmetry breaking can be diagnosed by long-range correlation of order parameter in the steady state:
\begin{equation}
\lim_{|i-j|\rightarrow\infty}\langle a_i^{\dagger}a_j\rangle\equiv\lim_{|i-j|\rightarrow\infty}\tr (\rho_{ss}a_i^{\dagger}a_j) \neq 0,
\label{eq:lro}
\end{equation}
which is invariant under weak symmetry operation. The other perspective for weak symmetry breaking is to look at the Liouvillian gap and steady state degeneracy in the thermodynamic limit. In the symmetry charge basis, a density matrix can be mapped to a pure state in the doubled space:

\begin{equation}
\rho = \sum_{jk}c_{jk}|j\rangle\langle k|\rightarrow |\rho\rrangle  = \sum_{jk}c_{jk}|j\rangle_L\otimes|k\rangle_R.
\label{eqn:double}
\end{equation}
In this new Hilbert space, weak symmetry implies the conservation of $N_L-N_R$, however, $N_L-N_R$ does not correspond to any physical observable.  For a general weak $U(1)$ symmetric Liouvillian $\mathcal{L}$, there always exists a steady state $\rho_{ss}^0$  in the sector $N_L-N_R = 0$, whether the symmetry is spontaneously broken or not (this is the steady state if we start from a pure state with fixed $N$). In the thermodynamic limit, if $\rho_{ss}^0$ is the only steady state (finite Liouvillian gap in sectors $N_L-N_R \neq 0$), we say the Liouvillian $\mathcal{L}$ is in the symmetric phase, since:
\begin{equation}
\mathcal{U}_{\text{w}}(\theta)[\rho_{ss}^0]\equiv e^{-iN\theta}\rho_{ss}^0e^{iN\theta} = \rho_{ss}^0,
\end{equation}
when the Liouvillian gap closes in sectors $N_L-N_R = n\neq 0$, there will be corresponding eigenstates $\rho_{ss}^n$ with 0 eigenvalue, i.e., $\mathcal{L}[\rho_{ss}^n] = 0$.
Then, with proper superposition:
\begin{equation}
\mathcal{U}_{\text{w}}(\theta)[\sum_n c_n\rho_{ss}^n] = \sum_ne^{-in\theta} c_n \rho_{ss}^n,
\end{equation}
the steady states now transform nontrivially under weak $U(1)$ symmetry due to steady state degeneracy. Consequently, weak $U(1)$ SSB is accompanied by the Liouvillian gap closing in sectors $N_L-N_R\neq 0$ in the thermodynamic limit.

In the weak $U(1)$ SSB phase, generally the steady states will be oscillatory. For a weak $U(1)$ SSB Liouvillian $\mathcal{L}$:
\begin{equation}
\mathcal{L}[\rho_{ss}^n] = 0,\quad n = 0,\pm 1,\pm 2\cdots
\end{equation}
if we add a term in the Hamiltonian that is proportional to the generator $N$, $\mathcal{L}\rightarrow\mathcal{L'} = \mathcal{L}-i\mu[N,]$:
\begin{equation}
\mathcal{L'}[\rho_{ss}^n] = -i\mu n\rho_{ss}^n,
\end{equation}
the eigenvalues now acquire an imaginary part $\mu n$, equally spaced on the imaginary axis. This is equivalent to going to the rotating frame $\rho\rightarrow \rho' = e^{-i\mu Nt}\rho e^{i\mu N t}$:
\begin{equation}
\frac{d}{dt}\rho' = \mathcal{L'}[\rho']\quad \Leftrightarrow \quad \frac{d}{dt}\rho = \mathcal{L}[\rho],
\end{equation}
where observables like the order parameter will now be oscillatory: $\langle a\rangle\rightarrow \langle a\rangle e^{-i\mu t}$. This time crystalline order resembles limit cycles and quantum synchronization in open systems \cite{Tindall_2020_Synchronization,Buca19nonstationary,Buca22Algebraic,Lee13synchronization,Lee14entanglement,Lee13unconventional}.

Now let us turn to strong symmetry and its possible spontaneous breaking. Strong $U(1)$ symmetry can be viewed as a special weak $U(1)$ symmetry, where not only $N_L-N_R$, but also $N_L,N_R$, are conserved in the doubled space, so that there are two $U(1)$ symmetries generated by $N_L-N_R$ and $\frac{1}{2}(N_L+N_R)$, respectively. SSB associated with generator $N_L-N_R$ is the same as weak symmetry breaking where we can also use long-range order Eq.~(\ref{eq:lro}) to characterize it. 

However, SSB associated with generator $\frac{1}{2}(N_L+N_R)$ is quite different. We argue that this symmetry is always spontaneously broken, for any parameter and in any dimension. For a strong $U(1)$ symmetric Liouvillian $\mathcal{L}$, there must exist steady states $\rho_{ss}^{nn}$ in every doubled space sector $N_L = N_R = n$, which has been noted in e.g. \cite{Buca_2012}. To understand it, we note that, since the total charge $N$ is conserved, in the physical space one will arrive at steady state $\rho_{ss}^{nn}$ starting from a pure state with fixed $N = n$. Consequently, for a strong $U(1)$ symmetric Liouvillian $\mathcal{L}$, we can always find steady states of the form $\sum_nc_n\rho_{ss}^{nn}$ that transforms non-trivially under strong symmetry operation Eq.~(\ref{eq:stro})(but not under weak symmetry operation Eq.~(\ref{eq:weak})):
\begin{align}
&\mathcal{U}_{\text{s}}(\theta)[\sum_n c_n\rho_{ss}^{nn}]\equiv U(\theta)\sum_n c_n\rho_{ss}^{nn} = \sum_ne^{-in\theta} c_n \rho_{ss}^{nn},\nonumber\\
&\mathcal{U}_{\text{w}}(\theta)[\sum_n c_n\rho_{ss}^{nn}] = \sum_n c_n \rho_{ss}^{nn},
\end{align}
which suggests that the symmetry generated by $\frac{1}{2}(N_L+N_R)$ is spontaneously broken. The weak symmetry generated by $N_R-N_L$ can remain unbroken, resulting in the strong-to-weak spontaneous symmetry breaking.

A natural question is whether we can use long-range order to describe strong-to-weak SSB. In the doubled space, the order parameter for strong-to-weak symmetry breaking is $a_{iL}\otimes a_{iR}$, which changes $\frac{1}{2}(N_L+N_R)$ but not $N_L-N_R$. Long-range order for strong-to-weak SSB should be described by:
\begin{equation}
\lim_{|i-j|\rightarrow\infty}\frac{\llangle\rho|a^{\dagger}_{iL}a_{jL}\otimes a_{iR}^{\dagger}a_{jR}|\rho\rrangle}{\llangle\rho|\rho\rrangle}\neq 0.
\end{equation}
In the physical space, this becomes:
\begin{equation}
\lim_{|i-j|\rightarrow\infty}\frac{\tr(\rho a_i^{\dagger}a_j\rho a_i a_j^{\dagger})}{\tr \rho^2}\neq 0,
\label{lro:swssb}
\end{equation}
which is the R\'enyi-2 correlator previously used to diagnose strong-to-weak SSB in mixed states \cite{lee2023quantum,lessa2024strong,sala2024spontaneous}. Since there are always steady states that explicitly break the strong $U(1)$ symmetry for a strong $U(1)$ symmetric Liouvillian, indicating SWSSB, we expect that the long-range order of Eq.~(\ref{lro:swssb}) should generally hold for the strongly symmetric steady states $\rho_{ss}^{nn}$. Later in Sec.~\ref{subsec:keldysh s}, we will describe how long-range order for strong-to-weak SSB arises in the field-theoretical description and how it persists in all dimensions.

It has been argued a lot of inequivalent imformation-theoretical quantities can be used to describe SWSSB \cite{lee2023quantum,lessa2024strong,sala2024spontaneous}, where different quantities can predict different transition thresholds as a strongly-symmetric channel acts on a symmetric state. Howver, for steady states we believe these quantities should yield the same results.

%This is known as the strong-to-weak symmetry breaking \cite{lee2023quantum,ma2023topological,lessa2024mixed,lessa2024strong,sala2024spontaneous}. 

Although the strong-to-weak SSB does not give rise to conventional long-range order linear in the density matrix, it has important physical consequences. In particular, the strong continuous symmetry together with translational invariance will render the Liouvillian $\mathcal{L}$ gapless. Physically, the Goldstone modes associated with broken symmetry generated by $\frac{1}{2}(N_L+N_R)$ correspond to diffusion of the conserved charge. Concrete examples of strong-to-weak symmetry breaking and the associated gapless spectrum will be given shortly in our discussion of models with strong $U(1)$ symmetry in Sec.~\ref{subsec:keldysh s}, and Sec.~\ref{sec:s-to-w}.

It is worth mentioning that similar discussion of emergent diffusive hydrodynamics in Brownian random time evolution through spontaneous symmetry breaking in the doubled space of the averaged Lindlabdian is discussed in \cite{PhysRevLett.131.220403}. %This can be understood as an enhanced Lieb-Schultz-Mattis (LSM)  theorem in open quantum systems \cite{LIEB1961407,PhysRevLett.84.1535,kawabata2023lieb}, where the gapless spectrum is protected by strong continuous symmetry and translational invariance at any filling. Noninteger filling number is not required in contrast to the closed system case \cite{PhysRevLett.84.1535}. 
\section{\label{sec:weak}Model with weak U(1) symmetry}
\subsection{The model}In this section, we focus on the case of weak $U(1)$ symmetry breaking. We consider a simple spin-$1/2$ model on a bipartite lattice with Liouvillian:
\begin{align}
&H = \sum_{\langle ij\rangle}J(\sigma_i^+\sigma_j^-+\sigma_i^-\sigma_j^+)+J_z\sigma_i^z\sigma_j^z,\nonumber\\
&\mathcal{L}_{\text{I}}[\rho]= -i[H,\rho]+\Gamma\sum_{i\in A}\mathcal{D}(\sigma_i^-)[\rho]+\Gamma\sum_{i\in B}\mathcal{D}(\sigma_i^+)[\rho],
\label{mod:weak}
\end{align}
where $\sigma^{\pm} = \frac{1}{2}(\sigma^x\pm i\sigma^y)$ and $\sigma^x,\sigma^y,\sigma^z$ are the Pauli operators. $\mathcal{D}(O)[\rho]$ is short for $2O\rho O^{\dagger}-\{O^{\dagger}O,\rho\}$.
The Hamiltonian part is the spin-$1/2$ XXZ model. The dissipation part describes onsite loss on the $A$ sublattice and onsite gain on the $B$ sublattice, with a uniform rate $\Gamma$.

$U(1)$ symmetry is generated by $N = \sum_i n_i = \sum_i(\sigma_i^z+1)/2$. This is equivalent to viewing the spins as hard-core bosons where a site $i$ is occupied when $\sigma_i^z = 1$ and unoccupied when $\sigma_i^z = -1$. Immediately we can see that there is no strong $U(1)$ symmetry ($N$ is not conserved) due to the presence of loss and gain. However, the Liouvillian $\mathcal{L}_{\text{I}}$ respects a weak $U(1)$ symmetry.
Possible SSB is diagnosed by long-range correlation $\langle\sigma_i^+\sigma_j^-\rangle$.
The competition between coherent Hamiltonian evolution and dissipation induces a symmetry-breaking transition when $J$ is increased from 0 across a critical value $J_c$. For simplicity, we set $J_z = 0$ and put the model on the square lattice in the following discussion.  
\subsection{SSB from gap closing}
In the trivial case where $J = 0$, the Liouvillian Eq.~(\ref{mod:weak}) is gapped and the unique steady state is:
\begin{equation}
\rho_0 = |\downarrow\downarrow\cdots\rangle\langle\downarrow\downarrow\cdots|_A\bigotimes|\uparrow\uparrow\cdots\rangle\langle\uparrow\uparrow\cdots|_B, 
\end{equation}
where there is no long-range order and thus no symmetry breaking.

When $J$ is increased from 0, we can treat the Hamiltonian $H$ as a perturbation. In the doubled space, the $J = 0$ steady state is in the sector $N_L-N_R = 0$. To diagnose symmetry breaking, we focus on the sectors where $N_L-N_R\neq 0$, for example, $N_L-N_R = 1$. At $J = 0$, in the long time, the important states (apart from $\rho_0$) are those with the largest eigenvalue $-\Gamma$.  These states are mapped to pure state $|i\rangle$($i$ is the site index):
\begin{equation}
\left \{
\begin{array}{ll}
    \sigma_i^+\rho_0,     & i\in A\\ 
    \rho_0\sigma_i^+,     & i\in B
\end{array}
\right. \rightarrow |i\rangle 
\end{equation}
They are the one magnon states. The Hamiltonian acts as the kinetic term for the magnons to hop between neighboring sites. To first order in perturbation theory, the effective Liouvillian $\mathcal{L}^{\text{eff}}_{\text{I}}$ in terms of $|i\rangle$'s is:
\begin{equation}
\mathcal{L}^{\text{eff}}_{\text{I}} = -\Gamma + iJ\sum_{\langle ij\rangle} |i\rangle\langle j|-|j\rangle\langle i|,\quad(i\in B, j\in A).
\label{equ:eff liou I}
\end{equation}
Detailed derivation is provided in Appendix \ref{app:perturbation}. For $\mathcal{L}_{\text{I}}^{\text{eff}}$, the $k = 0$ magnon (with $\pi/2$ phase difference between two sublattices) reduces the dissipative gap from $\Gamma$ to $\Gamma-2Jd$, where $d$ is the spatial dimension. Consequently, when $J$ is large, the magnons tend to condense at $k = 0$ and the Liouvillian gap closes, signaling weak $U(1)$ SSB as we discuss in Sec.~\ref{sec:sym}.
\subsection{\label{subsec:mean w}SSB from mean field}
A perhaps simpler way to see the symmetry-breaking process is through mean field theory. By mean field decoupling: $\rho = \bigotimes_i\rho_i,\,\langle\sigma_i\sigma_j\rangle = \langle\sigma_i\rangle\langle\sigma_j\rangle$,
together with the Heisenberg equation for operators:
\begin{equation}
\partial_tO = -i[O,H]+\sum_{\mu}\gamma_{\mu}\bigg(2L_{\mu}^{\dagger}OL_{\mu}-\{L_{\mu}^{\dagger}L_{\mu},O\}\bigg),
\label{eq:heisen}
\end{equation}
we arrive at the mean field equations:
\begin{align}
&\frac{d}{dt}\sigma_{i\in A}^+ = -\Gamma \sigma_{i\in A}^+-iJ\sum_{\langle ij\rangle}\sigma^z_{i\in A}\sigma_{j\in B}^+,\nonumber\\
&\frac{d}{dt}\sigma_{i\in A}^z = -2\Gamma(\sigma_{i\in A}^z+1)+2iJ\sum_{\langle ij\rangle}(\sigma_{i\in A}^-\sigma_{j\in B}^+-\sigma_{i\in A}^+\sigma_{j\in B}^-),\nonumber\\
&\frac{d}{dt}\sigma_{i\in B}^+ = -\Gamma \sigma_{i\in B}^+-iJ\sum_{\langle ij\rangle}\sigma^z_{i\in B}\sigma_{j\in A}^+,\nonumber\\
&\frac{d}{dt}\sigma_{i\in B}^z = 2\Gamma(1-\sigma_{i\in B}^z)+2iJ\sum_{\langle ij\rangle}(\sigma_{i\in B}^-\sigma_{j\in A}^+-\sigma_{i\in B}^+\sigma_{j\in A}^-).
\label{eq:mean w}
\end{align}
Here each variable represents the expectation value of the corresponding operator. When $2Jd\le\Gamma$, the mean field equations have a trivial symmetric steady state with $\sigma_A^z = -1, \sigma_B^z = 1,\sigma_A^+ = \sigma_B^+ = 0$. When $2Jd>\Gamma$, we arrive at symmetry breaking steady state with a nonzero order parameter:
\begin{align}
&\sigma_A^z = -\frac{\Gamma}{2Jd},\quad\sigma_B^z = \frac{\Gamma}{2Jd},\nonumber\\
&\sigma_A^+ = i\sigma_B^+,\quad|\sigma_A^+|^2 = \frac{1}{2}\frac{\Gamma}{2Jd}(1-\frac{\Gamma}{2Jd}),
\label{eq:SSB w}
\end{align}
where on each sublattice the order parameter $\sigma_i^+$ takes a uniform value and there is $\pi/2$ phase difference between order parameters on the two sublattices.

The mean field theory result is consistent with previous perturbation analysis where we find that magnon condensation leads to Liouvilian gap closing in non-diagonal sectors $N_L-N_R\neq 0$ and there is $\pi/2$ phase difference between the condensed magnon on two sublattices.
\subsection{\label{subsec:keldysh w}Field-theoretical analysis and lower critical dimension}
To further study the model and its weak $U(1)$ symmetry-breaking transition, we employ Keldysh field theory to see whether long-range order survives fluctuations beyond mean field.

The Keldysh path integral is formally expressed as:
\begin{equation}
\int\mathcal{D}[\psi_+,\psi_-]\,e^{iS[\psi_+,\psi_-]},
\label{eq:kel formalism}
\end{equation}
where $\psi_+(x,t), \psi_-(x,t)$ are defined on the forward (backward) branch of a closed time contour. It is often convenient to perform the Keldysh rotation $\psi_c = \frac{1}{\sqrt{2}}(\psi_++\psi_-),\,\psi_q = \frac{1}{\sqrt{2}}(\psi_+-\psi_-)$, where $\psi_{c/q}$ are called classical (quantum) fields. At the saddle point
\begin{equation}
\frac{\delta S}{\delta \psi_c} = 0,\quad \frac{\delta S}{\delta \psi_q} = 0.
\label{eq:saddle}
\end{equation}
Typically $\langle\psi_q\rangle = 0$, and its fluctuations are gapped. 

Models similar to Eq.~(\ref{mod:weak}) with weak $U(1)$ symmetries have been studied in \cite{PhysRevB.93.014307} using Keldysh formalism. Close to symmetry breaking transition point $J = J_c$, mean field theory gives: $\sigma_{i\in A}\approx -1,\,\sigma_{i\in B}\approx 1$. Consequently, we choose $|\downarrow_A\uparrow_B\rangle$ as the reference state, and use the large-spin Holstein–Primakoff transformation to Eq.~(\ref{mod:weak}) to a bosonic model with Keldysh action $S_1[\psi_+,\psi_-]$. The details of large-spin expansion is irrelevant. For the long wavelength universal behaviors near the critical point $J = J_c$, only symmetry matters. The action $S_1$ respects a global $U(1)$ symmetry:
\begin{equation}
S_1[\psi_+,\psi_-] = S_1[\psi_+e^{i\phi},\psi_-e^{i\phi}],
\label{eq:sym s1}
\end{equation}
which is the manifestation of weak $U(1)$ symmetry in Eq.~(\ref{mod:weak}). (A subtle point is that Eq.~(\ref{mod:weak}) is not fully translational invariant as different dissipators are imposed on different sublattices, so a transformation is made to restore the full translational invariance). $S_1$ is in the same universality class as the driven-dissipative BEC studied in \cite{Sieberer_2016,PhysRevB.89.134310,PhysRevX.5.011017,PhysRevLett.110.195301}, and we review the results here. For $d \geq 3$, the Keldysh action $S_1$ is of the form:
\begin{align}
S_1 = &\int dtd^dx\,\bigg\{\psi_q^*[i\partial_t+(K_c-iK_d)\nabla^2-r_c+ir_d]\psi_c + \text{c.c.}\nonumber\\
&-[(u_c-iu_d)\psi_q^*\psi_c^*\psi_c^2+\text{c.c.}]+2i\gamma\psi_q^*\psi_q\bigg\},
\label{eq:action weak}
\end{align}
where $K_c, K_d, r_c, r_d, u_c, u_d,\gamma$ are phenomenological parameters arising from different terms in the microscopic model. All terms allowed by weak $U(1)$ symmetry are considered in $S_1$, with terms irrelevant in the RG sense dropped. Close to the critical point $r_d = 0$ ($r_c$ is unimportant because we can always go to a rotating frame $\psi_c\rightarrow \psi_ce^{i\omega t}$ and shift $r_c$ to 0), the scaling dimension is: $[\nabla] = 1,\,[\partial_t] = 2,\,[\psi_c] = (d-2)/2,\,[\psi_q] = (d+2)/2$.
Terms containing more than two $\psi_q's$ are irrelevant, and thus $S_1$ Eq.~(\ref{eq:action weak}) is quadratic in $\psi_q$.

Since $\psi_q$ is also gapped in $S_1$, we can integrate it out \cite{Sieberer_2016} and arrive at the following Langevin equation for the order parameter $\psi_c$:
\begin{equation}
i\partial_t\psi_c = [-(K_c-iK_d)\nabla^2+r_c-ir_d+(u_c-iu_d)|\psi_c|^2]\psi_c+\xi,
\label{eq:langevin weak}
\end{equation}
where $\xi$ is Guassian white noise with $\langle\xi(x,t)\rangle = 0$ and $\langle\xi(x,t)\xi^*(x',t')\rangle = 2\gamma \delta(t-t')\delta(x-x')$. 

The Langevin equation Eq.~(\ref{eq:langevin weak}) captures the long-distance and long-time physics. Symmetry breaking transition is controlled by $r_d\sim\Gamma-2dJ$. When $r_d>0(J<J_c)$, the system is in the symmetric phase with $\langle\psi_c\rangle = 0$. When $r_d<0 (J> J_c)$, the system is in the symmetry-broken phase with long-range order in $\psi_c$, and the long-time dynamics is described by the phase fluctuation of $\psi_c = \sqrt{\rho_0}e^{i\theta_c}$:
\begin{equation}
\partial_t\theta_c = D\nabla^2\theta_c + \frac{\lambda}{2}(\nabla\theta_c)^2 + \eta(x,t),
\label{eq:kpz}
\end{equation}
which is the famous KPZ equation \cite{HALPINHEALY1995215,PhysRevLett.56.889}, where $\rho_0 = -r_d/u_d,\,D = K_cu_c/u_d,\,\lambda = -2K_c$. $\eta(x,t)$ is Gaussian white noise satisfying $\langle\eta(x,t)\rangle = 0,\,\langle\eta(x,t)\eta(x',t')\rangle = 2\Delta\delta(x-x')(t-t')$, where $\Delta = (\gamma+2u_d\rho_0)(u_c^2+u_d^2)/(2\rho_0u_d^2)$.

In dimension $d\geq 3$, the nonlinear term $(\nabla\theta_c)^2$ is irrelevant. Phase fluctuations do not destroy long-range order, with dynamical critical exponent $z = 2$ and Goldstone modes corresponding to fluctuations of the order parameter (spin waves). For $d<3$, it can be shown \cite{Sieberer_2016} that the long-time dynamics is still governed by Eq.~(\ref{eq:kpz}), and phase fluctuations completely destroy long-range order in the steady state. Notably, due to the nonlinear term $(\nabla\theta_c)^2$, there is no quasi-long-range order in 2D \cite{Sieberer_2016,PhysRevX.5.011017}.

Generally, the weak $U(1)$ symmetry breaking with long-range ordered steady states can occur in $d\geq 3$.   %(Eq.~(\ref{eq:lro})).

\section{\label{sec:strong} Model with strong U(1) symmetry}
\subsection{The model}
In this section, we consider a model with strong $U(1)$ symmetry. It is described by Liouvillian $\mathcal{L}_{\text{II}}$:
\begin{align}
H = &\sum_{\langle ij\rangle}J(\sigma_i^+\sigma_j^-+\sigma_i^-\sigma_j^+)+J_z\sigma_i^z\sigma_j^z,\nonumber\\
\mathcal{L}_{\text{II}}[\rho]= & -i[H,\rho] +\Gamma\sum_{\langle ij\rangle}\mathcal{D}(\sigma_{j\in B}^+\sigma_{i\in A}^-)[\rho]\nonumber\\
& + \Gamma_z\sum_i\mathcal{D}(\sigma_i^z)[\rho],
\label{mod:strong}
\end{align}
The strong $U(1)$ symmetric Liouvillian $\mathcal{L}_{\text{II}}$ share the same bipartite lattice structure and Hamiltonian with $\mathcal{L}_{\text{I}}$ Eq.~(\ref{mod:weak}), however, the dissipation parts are markedly different. Instead of raising and lowering spin (creating and annihilating hard-core boson) on each site, the jump operators here incoherently move hard-core bosons  between nearest sites (from $A$ sublattice to $B$ sublattice) with rate $\Gamma$. Local dephasing terms represented by jump operators $\sigma_i^z$ with strength $\Gamma_z$ are also included.

$\mathcal{L}_{\text{II}}$ respects a strong $U(1)$ symmetry, which means the total particle number $N = \sum_i(\sigma_i^z+1)/2$ is conserved: $\frac{d}{dt}\langle N \rangle = 0$. This is the major difference between model I and model II. Later we will find out that the conservation of $N$ has profound consequences. 

As discussed in Sec.~\ref{sec:sym}, for a strong $U(1)$ symmetric Liouvillian, there are actually two $U(1)$ symmetries generated by $N_L+N_R$ and $N_L-N_R$ in doubled space. The $U(1)$ symmetry generated by $N_L-N_R$ is similar to the weak symmetry, and its symmetry breaking can be characterized by a long-range order $\lim_{|i-j|\rightarrow\infty}\langle \sigma_i^+\sigma_j^-\rangle\neq 0$. Furthermore, we find that for $\mathcal{L}_{\text{II}}$ complete breaking of the strong $U(1)$ symmetry and long-range order occur for a certain range of fillings. On the other hand, the $U(1)$ symmetry generated by $N_L+N_R$ always spontaneously breaks. It can be understood as the  strong-to-weak symmetry breaking, which
leads to gapless diffusion modes of the conserved charge. We first focus on weak $U(1)$ SSB of model II and leave the discussion of strong-to-weak symmetry breaking to Sec.~\ref{subsec:keldysh s}.
%As discussed in Sec.~\ref{sec:sym}, for a strong $U(1)$ symmetric Liouvillian, there are actually two $U(1)$ symmetries generated by $N_L+N_R$ and $N_L-N_R$ in doubled space. The $U(1)$ symmetry generated by $N_L-N_R$ is similar to the weak symmetry, and its symmetry breaking can be characterized by a long-range order $\lim_{|i-j|\rightarrow\infty}\langle \sigma_i^+\sigma_j^-\rangle\neq 0$. On the other hand, the $U(1)$ symmetry generated by $N_L+N_R$ always spontaneously breaks, which can be understood as strong-to-weak
%symmetry breaking. We find that for $\mathcal{L}_{\text{II}}$, complete breaking of the strong $U(1)$ symmetryand long-range order only occur for certain range of fillings, and strong-to-weak symmetry breaking leads to gapless diffusion modes of the conserved charge. We first focus on weak $U(1)$ SSB of model II and leave the discussion of strong-to-weak symmetry breaking to Sec.~\ref{subsec:keldysh s}.

\subsection{\label{subsec:mean s} Mean field analysis}
We consider weak $U(1)$ symmetry breaking of model II. For simplicity, we consider the model on the square lattice and set $J_z = \Gamma_z = 0$ for the time being. By performing mean field approximation as in Sec.~\ref{subsec:mean w}, we arrive at equations of motion:
\begin{align}
\frac{d}{dt}\sigma_{i\in A}^+ = &\sum_{\langle ij\rangle}-\frac{\Gamma}{2}  \sigma_{i\in A}^+(1-\sigma_{j\in B}^z)-iJ\sigma^z_{i\in A}\sigma_{j\in B}^+,\nonumber\\
\frac{d}{dt}\sigma_{i\in A}^z = &\sum_{\langle ij\rangle}\,\Gamma (\sigma_{i\in A}^z+1)(\sigma_{j\in B}^z-1)\nonumber\\
&+2iJ(\sigma_{i\in A}^-\sigma_{j\in B}^+-\sigma_{i\in A}^+\sigma_{j\in B}^-),\nonumber\\
\frac{d}{dt}\sigma_{i\in B}^+ = &\sum_{\langle ij\rangle}-\frac{\Gamma}{2} \sigma_{i\in B}^+(1+\sigma_{j\in A}^z) -iJ\sigma^z_{i\in B}\sigma_{j\in A}^+,\nonumber\\
\frac{d}{dt}\sigma_{i\in B}^z = &\sum_{\langle ij\rangle}\,\Gamma  (1-\sigma_{i\in B}^z)(1+\sigma_{j\in A}^z)\nonumber\\
& +2iJ(\sigma_{i\in B}^-\sigma_{j\in A}^+-\sigma_{i\in B}^+\sigma_{j\in A}^-);
\label{eq:mean s}
\end{align}
where each variable represents its expectation value. The conservation of $N$ manifests as $\frac{d}{dt} \sum_i\sigma_i^z = 0$. We define the filling factor 
as $n = N/L^d = \sum_i\frac{1}{2}(\sigma_i^z+1)/L^d$($L^d$ is the number of total sites). By definition, $0\leq n\leq 1$. The model respects a particle-hole symmetry: $\sigma_i^+\leftrightarrow \sigma_i^-,\sigma_i^z\leftrightarrow -\sigma_i^z,A\leftrightarrow B$. Consequently, we consider only $0\leq n\leq 1/2$.

For $0\leq n\leq 1/4$, the mean field equations of motion  Eq.~(\ref{eq:mean s})  give no weak $U(1)$ symmetry-broken steady states. Since the dissipation we consider in $\mathcal{L}_{\text{II}}$ takes hard-core bosons from the $A$ sublattice to the $B$ sublattice, when the filling factor is small, the hard-core bosons prefer to stay on the $B$ sublattice. 

One possible stable steady state satisfies $\sigma_A^+ = \sigma_B^+ = 0,\,\sigma_A^z = -1$. The expectation value of order parameter $\sigma^+$ is 0, $\sigma_A^z = -1$ implies no hard core bosons on A sublattice. The value of $\sigma_{i\in B}^z$ is uniform on the $B$ sublattice and determined by filling factor $n$. This is by definition a symmetric steady state. When $n$ is tuned above $1/4$, it becomes unstable (infinitesimal deviation from this steady-state solution will drive the system toward a symmetry-breaking state).

\begin{figure}[t]
    \centering
    \includegraphics[width=\linewidth]{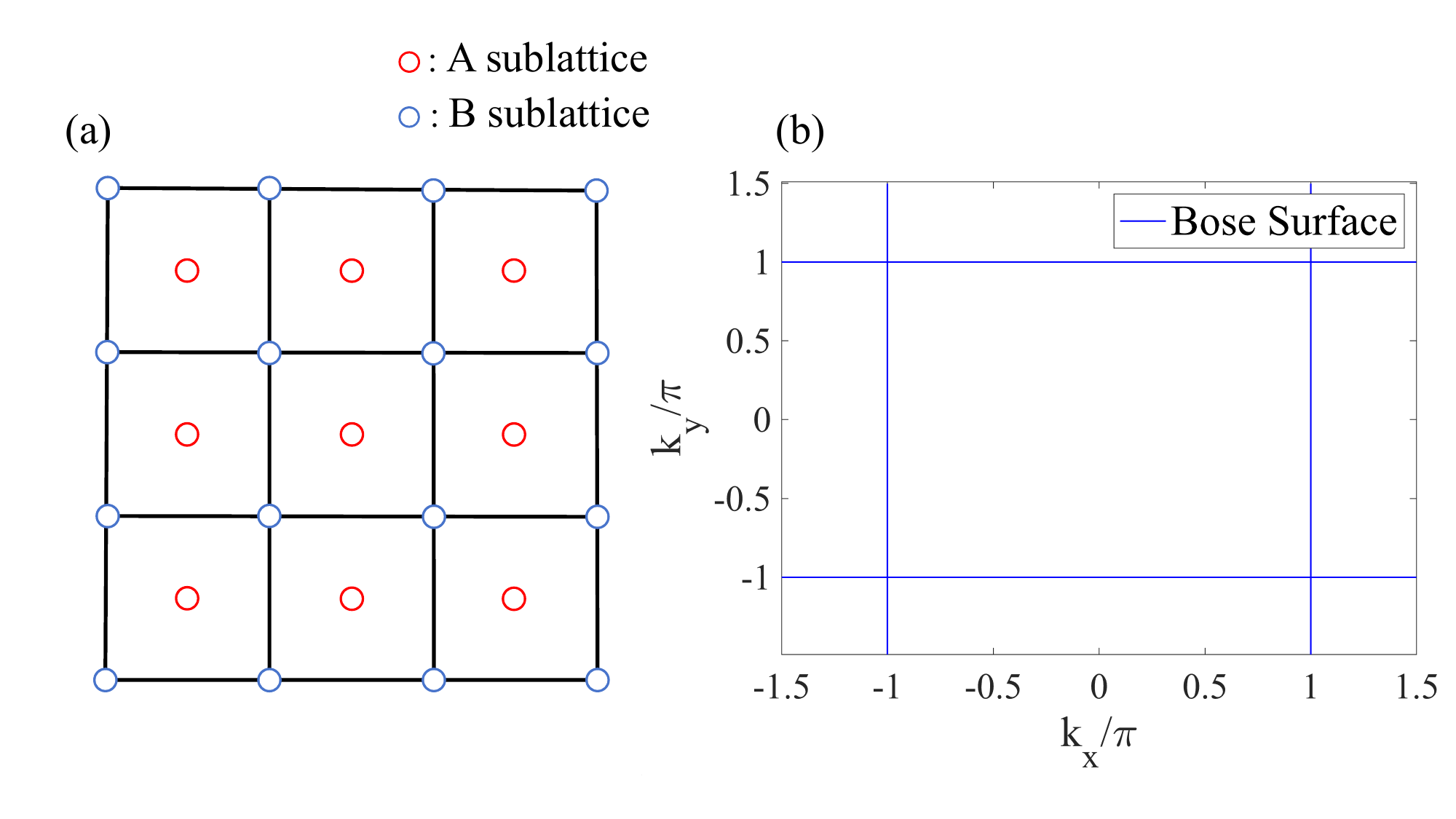}
    \caption{(a) We consider square lattice case, where the $B$ sublattice is still a square lattice. (b) the Bose surface on the $B$ sublattice.}
    \label{fig:bs}
\end{figure}

The other possible steady state for $0\leq n\leq 1/4$ satisfies:
\begin{equation}
\sigma_A^z = -1,\,\sigma_A^+ = 0;\quad\sum_{\langle ij\rangle}\sigma^+_{j\in B} = 0,\,\forall i\in A;
\label{eq:bose surface}
\end{equation}
where there is still no hard-core bosons on the $A$ sublattice. However, on the $B$ sublattice there can be nonzero order parameter $\sigma_{i\in B}^+$ as long as it satisfies Eq.~(\ref{eq:bose surface}): it is possible only with momentum $k$ on a certain Bose surface\cite{subir02bs,Varney11bs,Wang2020pattern}:
\begin{equation}
\sigma_{r\in B}^+ = \sum_k \sigma_k^+ e^{ik\cdot r},\quad \prod_{i = 1}^d\cos k_i/2 = 0.
\label{eq:bose surface 2}
\end{equation}
The Bose surface is illustrated in Fig.~\ref{fig:bs}. However, the Bose surface steady states  Eq.~(\ref{eq:bose surface}) are unstable against dephasing. When $\Gamma_z > 0$, it is no longer a steady state. Nonzero order parameter now vanishes: $\forall i, \sigma_i^+ = 0$.

For $1/4\leq n\leq 1/2$, mean field equations of motion lead to symmetry breaking steady state, where $\sigma_i^{\pm},\sigma_i^z$ takes uniform value on each sublattice and satisfies:
\begin{align}
&\Gamma^2(1-\sigma_B^z)(1+\sigma_A^z) = -4J^2\sigma_A^z\sigma_B^z,\nonumber\\
&\Gamma(1-\sigma_B^z)\sigma_A^+ = -2\,iJ\sigma_A^z\sigma_B^+,\nonumber\\
&2|\sigma_A^+|^2 = -\sigma_A^z(1+\sigma_A^z);
\label{eq:SSB s}
\end{align}
and the filling factor is $n = (\sigma_A^z+\sigma_B^z+2)/4$. (The reason that SSB steady state is possible only for $1/4\leq n\leq3/4$ is that the SSB steady state Eq.~(\ref{eq:SSB s}) is possible only when $\sigma_A^z\leq 0, \sigma_B^z\geq 0$). Exactly at $n = 1/2$, the steady state satisfies (to first order in $J/\Gamma$): 
\begin{align}
&\sigma_A^z = -1 + \frac{2J}{\Gamma},\quad \sigma_B^z = 1 - \frac{2J}{\Gamma};\nonumber\\
&\sigma_B^+ = i\sigma_A^+,\quad |\sigma_A^+| = \sqrt{\frac{J}{\Gamma}}.
\label{eq:mf s n05}
\end{align}
Similar to the SSB steady state of $\mathcal{L}_{\text{I}}$ Eq.~(\ref{eq:SSB w}), there is also a $\frac{\pi}{2}$ phase difference between order parameter on two sublattices. When there is no dephasing, infinitesimal $J$ will drive the system into the symmetry-broken phase. Generally, when $\Gamma_z>0$, there will be a phase transition: $J$ has to be larger than a critical value $J_c$ for the system to be in the symmetry-broken phase.

\subsection{Large spin analysis for n = 0 and n = 1/2}
One interesting result from mean field theory is that the model $\mathcal{L}_{\text{II}}$ goes from symmetric phase to symmetry-broken phase when filling factor $n$ is increased from 0 to $\frac{1}{2}$, this is a unique phenomenon for strong symmetric Liouvillian since particle number conservation requires strong symmetry. To better understand the transition, we first focus on the cases $n = 0$ and $n = 1/2$.

We still consider $J_z = \Gamma_z = 0$. At exactly $n = 0$, the steady state is trivial: $\rho_{ss} = \rho_1 = |\downarrow\downarrow\cdots\rangle\langle\downarrow\downarrow\cdots|$. When the system is doped a little away from $n = 0$, we believe the long-time dynamics will only involve states close to $\rho_1$, which allows us to perform large-$S$ Holstein–Primakoff expansion to the spins (by taking $|\downarrow\downarrow\cdots\rangle$ as the reference state):
\begin{equation}
S^z_i = -S + b_i^{\dagger}b_i,\quad S_i^{+} = b_i^{\dagger}\sqrt{2S-b_i^{\dagger}b_i};
\label{eq:largesn0}
\end{equation}
to zeroth order in $1/S$, we have: $\sigma_i^+ = S_i^+ = b_i^{\dagger},\sigma_i^z = 2S_i^z = 2b_i^{\dagger}b_i-1$. This is equivalent to relaxing the hard-core boson constraint $n_i\leq 1$. Consequently, near $n = 0$, $\mathcal{L}_{\text{II}}$ is approximately described by the Liouvillian:
\begin{equation}
\mathcal{L}^{\text{eff}}_{n = 0}[\rho] = -i[J\sum_{\langle ij\rangle}(b_i^{\dagger}b_j+b_j^{\dagger} b_i),\rho]+\Gamma\sum_{\langle ij \rangle}\mathcal{D}(b_{j\in B}^{\dagger}b_{i\in A})[\rho],
\label{eq:eff liou 0}
\end{equation}
for $\mathcal{L}^{\text{eff}}_{n = 0}$, there is no weak $U(1)$ SSB phase (mean field equations do not admit stable SSB steady state). However, there are exact dark states $|\psi\rangle$'s that satisfy:
\begin{equation}
\sum_{\langle ij\rangle}(b_i^{\dagger}b_j+b_j^{\dagger} b_i)|\psi\rangle = 0, \quad \forall\langle ij\rangle,\,b_{j\in B}^{\dagger}b_{i\in A}|\psi\rangle = 0,
\end{equation}
which makes $|\psi\rangle\langle\psi'|$ steady states of $\mathcal{L}^{\text{eff}}_{n = 0}$ Eq.~(\ref{eq:eff liou 0}): $\mathcal{L}^{\text{eff}}_{n = 0}[|\psi\rangle\langle\psi'|] = 0$. These steady states  correspond to the Bose surface steady states in our mean field theory analysis Eq.~(\ref{eq:bose surface}) and (\ref{eq:bose surface 2}), for that $|\psi\rangle$'s can be expressed as:
\begin{align}
&|\psi\rangle = b_{k_1}^{\dagger}b_{k_2}^{\dagger}\cdots b^{\dagger}_{k_n}|\Omega\rangle,\nonumber\\
&b_k^{\dagger} = \frac{1}{\sqrt{L^d/2}}\sum_{r\in B} b_r^{\dagger}e^{ikr},\quad n = 0,1,2\cdots
\label{eq:bose surface 3}
\end{align}
where $|\Omega\rangle$ is the vacuum for bosons and $k's$ are on the Bose surface Eq.~(\ref{eq:bose surface 2}). As in the mean field steady state, $A$ sites are empty, and bosons staying on the $B$ sites are dissipationless only with momentum on the Bose surface. Again, the Bose surface excitations are unstable against dephasing.

Now we turn to $n \approx 1/2$. By examining the mean field steady state Eq.~(\ref{eq:mf s n05}), when $J/\Gamma$ is small, we take the state $|\downarrow_{A}\uparrow_{B}\rangle$ as the vacuum for large $S$ expansion:
\begin{equation}
\left \{
\begin{array}{ll}
    S^z_i = -S + b_i^{\dagger}b_i,\quad S_i^{+} = b_i^{\dagger}\sqrt{2S-b_i^{\dagger}b_i},    & i\in A;\\ 
    S^z_i = S - b_i^{\dagger}b_i,\quad S_i^{+} = \sqrt{2S-b_i^{\dagger}b_i}\,b_i,     & i\in B;
\end{array}
\right.
\label{eq:largesn05}
\end{equation}
Similarly, to zeroth order in $1/S$, the effective Liouvillian is:
\begin{equation}
\mathcal{L}^{\text{eff}}_{n = \frac{1}{2}}[\rho] = -i[J\sum_{\langle ij\rangle}(b_i^{\dagger}b_j^{\dagger}+b_jb_i),\rho]+\Gamma\sum_{\langle ij \rangle}\mathcal{D}(b_{i}b_{j})[\rho],
\label{eq:eff liou 05}
\end{equation}
where $b_i^{\dagger}$ creates a particle on the $A$ sublattice and a hole on the $B$ sublattice.
Strong $U(1)$ symmetry implies the conservation of $\sum_{i\in A}b_i^{\dagger}b_i-\sum_{i\in B}b_i^{\dagger}b_i$. Quite different from $\mathcal{L}^{\text{eff}}_{n = 0}$, in $\mathcal{L}^{\text{eff}}_{n = 1/2}$ bosons are created, annihilated, and damped in pairs, leading to condensation and SSB \cite{Wang2020pattern}.  Within mean field theory, $\mathcal{L}^{\text{eff}}_{n = 1/2}$ indeed admits stable SSB steady state with uniform order parameter $\langle b_i\rangle$ on the A/B sublattice:
\begin{equation}
|b_A| = \sqrt{n_A},\quad\Gamma n_Bb_A = -iJb_B^{\dagger},\quad \Gamma^2n_An_B = J^2,
\end{equation}
where each variable denotes its expectation value.
Later in Sec.~\ref{subsec:keldysh s} we will use Keldysh field theory to further study $\mathcal{L}^{\text{eff}}_{n = 1/2}$, identifying lower critical dimension for weak $U(1)$ SSB and emergent gapless diffusion modes from strong-to-weak symmetry breaking.

\subsection{Transition from n = 0 to n = 1/2}
With a better understanding of $n = 0$ and $n = 1/2$, we now focus on the transition from $n = 0$ to $n = 1/2$. For $\mathcal{L}_{\text{II}}$ Eq.~(\ref{mod:strong}), the Hamiltonian $H$ is now treated  perturbatively (again we set $J_z = \Gamma_z = 0$ for simplicity).

At $J = 0$, the Liouvillian $\mathcal{L}_{\text{II}}$ is purely dissipative. There are numerous dark states on which dissipation has no effect:
\begin{equation}
\forall \langle ij\rangle,\quad \sigma_{j\in B}^{\dagger}\sigma_{i\in A}|\psi_d\rangle = 0,
\label{eq:dark}
\end{equation}
for example, states with hard-core bosons (spin up) only on the $B$ sublattice. The $J = 0$ steady states are of the form $|\psi_d\rangle\langle\psi_d'|$ or their linear superposition. As a result, there is massive degeneracy in each symmetry sector. When the Hamiltonian $H$ is added perturbatively, it mixes the degenerate steady states by connecting different configurations in $|\psi\rangle_d$'s, and through second-order perturbation picks certain superposition of $|\psi_d\rangle\langle\psi_d'|$ as the new steady state.

In $|\psi_d\rangle$'s, at a small filling factor, the hard-core bosons tend to occupy $B$ sites. The simplest case is to consider the symmetry sector $N_L = 1, N_R = 0$. $J \neq 0$ steady state is simple:
\begin{equation}
\rho_{ss} = \frac{1}{\sqrt{L^d/2}}\sum_{r\in B}\sigma_r^{+}e^{ikr}\,|\downarrow\downarrow\cdots\rangle\langle\downarrow\downarrow\cdots|,
\label{eq:magnon n0}
\end{equation}
where $k$ is on the Bose surface Eq.~(\ref{eq:bose surface 2}). When there are few bosons, $A$ sites are rarely occupied, and the bosons prefer to form magnons on the $B$ sublattice with momentum $k$ on the Bose surface, as in Eq.~(\ref{eq:bose surface 3}).

\begin{figure}[t]
    \centering
    \includegraphics[width=\linewidth]{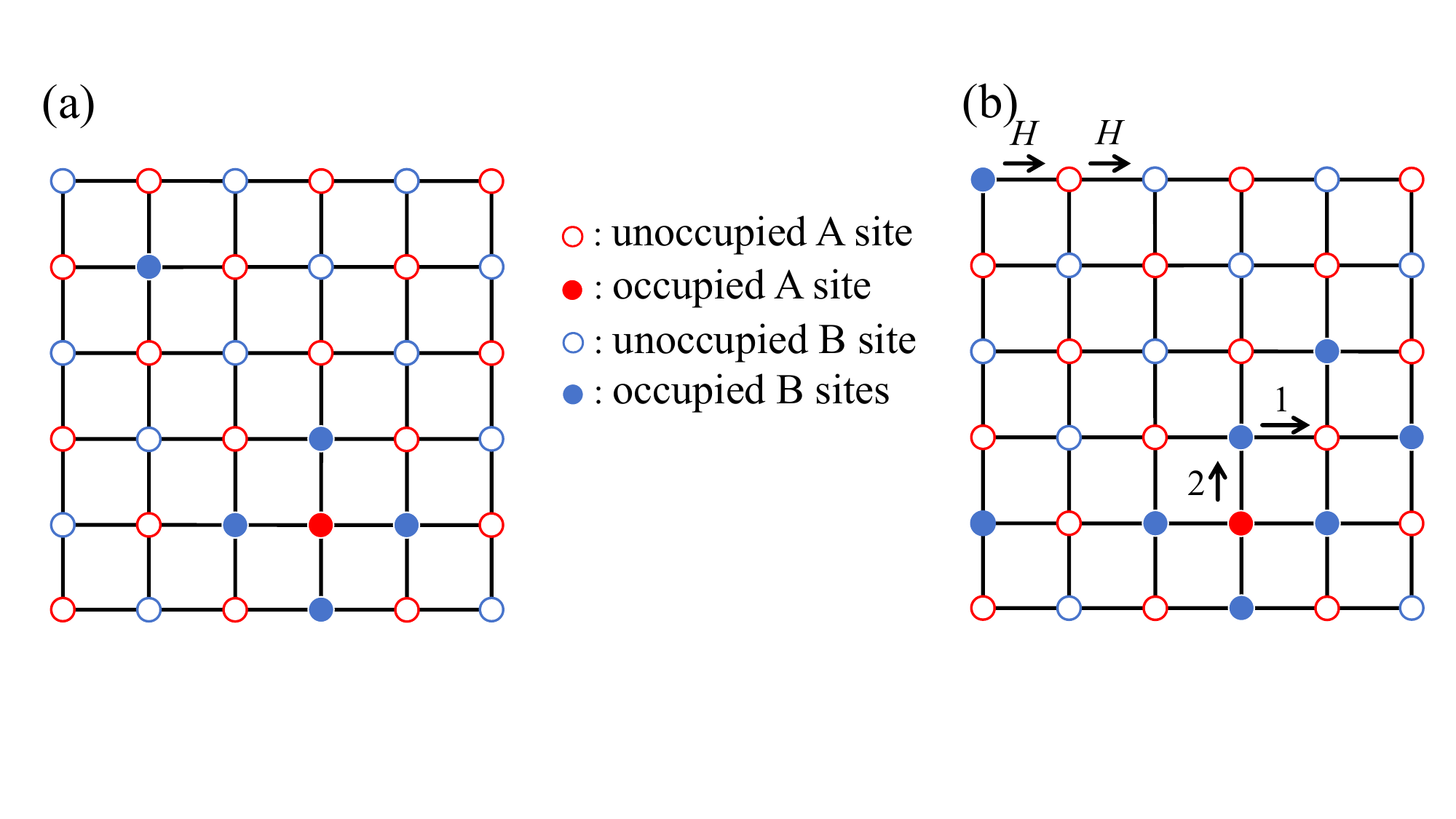}
    \caption{(a) In the dark states $|\psi_d\rangle$'s of Eq.~(\ref{eq:dark}), $B$ sites are freely occupiable, whereas an $A$ site can only be occupied if its neighboring $B$ sites are also occupied. (b) Perturbatively, the Hamiltonian connects different $|\psi_d\rangle$ configurations. Bosons can hop between the $B$ sublattice sites via second-order perturbation of $H$. On the $A$ sublattice, bosons can only move if the neighboring $B$ sites form a connected cluster of occupation.}
    \label{fig:hopping}
\end{figure}

To go into the symmetry-broken phase, the hard-core bosons should be able to explore the whole lattice and condense at $k = 0$. However, in $|\psi_d\rangle$'s, an $A$ site can be occupied only if the neighboring $B$ sites are all occupied, and an $A$ site boson can move through second-order perturbation of $H$ only if it is in a connected cluster of occupied $B$ sites (These are illustrated in Fig.~(\ref{fig:hopping})). Consequently, to achieve long-range order, occupied $B$ sites need to form a large enough connected cluster across the lattice in order for the $A$ site bosons to move around and create long-range coherence. This is similar to the percolation problem \cite{stauffer2018introduction}, where a large enough cluster across the lattice occurs only when filling factor 
 $n$ is larger than a critical value $n_c$. 

When $n$ is large enough, in the $|\psi_d\rangle$'s, occupied $B$ sites form a large cluster across the lattice. Especially, near $n = 1/2$, $\rho_0 = |\downarrow_A\uparrow_B\rangle\langle\downarrow_A\uparrow_B|$ is a $J = 0$ steady state. By doping particles on the $A$ sublattice or holes on the $B$ sublattice, $\sigma_{i_1\in A}^{+}\sigma_{i_2\in A}^{+}\cdots\rho_0$ and $\sigma_{i_1\in B}^-\sigma_{i_2\in B}^-\cdots\rho_0$ are still $J = 0$ steady states. When $J$ is increased from 0, in contrast to $n = 0$ case Eq.~(\ref{eq:magnon n0}), the perturbative action of $H$ will favor  $k = 0$ superposition of these doped states. More precisely, near $n = 1/2$, the model is described by the effective Liouvillian $\mathcal{L}^{\text{eff}}_{n = 1/2}$ Eq.~(\ref{eq:eff liou 05}), where $b_i^{\dagger}$ creates particles on the $A$ sublattice or holes on the $B$ sublattice. Pairs of $A$ particle and $B$ hole are created and annihilated coherently, which leads to SSB and uniform ($k = 0$) order parameter $\langle b_i\rangle$ on each sublattice.

Mean field analysis in Sec.~\ref{subsec:mean s} predicts a transition at $n_c = 1/4$.  However, with the above percolation analogy, the actual transition filling factor may not be $1/4$. Generally, it will depend on dimensionality and lattice geometry.

\subsection{\label{subsec:keldysh s}Field-theoretical analysis: lower critical dimension and emergent hydrodynamics}
For strong $U(1)$ symmetric $\mathcal{L}_{\text{II}}$ Eq.~(\ref{mod:strong}), long-range order is possible only for certain fillings. Here we focus on filling factor $n$ near $1/2$, where it is mapped to a bosonic model by large spin expansion Eq.~(\ref{eq:largesn05}). To zeroth order, the corresponding effective Liouvillian is $\mathcal{L}^{\text{eff}}_{n = 1/2}$ Eq.~(\ref{eq:eff liou 05}). 

For $\mathcal{L}_{\text{II}}$, following the same procedure in Sec.~\ref{subsec:keldysh w}, we should arrive at a keldysh action $S_2[\psi_+,\psi_-]$. Since the microscopic model Eq.~(\ref{mod:strong}) respects strong $U(1)$ symmetry, the action $S_2$ satisfy:
\begin{equation}
S_2[\psi_+,\psi_-] = S_2[\psi_+e^{i\phi_+},\psi_-e^{i\phi_-}].
\label{eq:sym s2}
\end{equation}
There are two global $U(1)$ symmetries corresponding to the $U(1)$ rotation of $\psi_+$ and $\psi_-$. In the doubled space, these two $U(1)$ symmetries are generated by $N_L$ and $N_R$. 

For the saddle point Eq.~(\ref{eq:saddle}) of $S_2$, $\psi_q = 0$ is still a solution. However, the term of the form $|\psi_q|^2$ is not invariant under the symmetry operation Eq.~(\ref{eq:sym s2}), which means $\psi_q$ can not be integrated out in the same way as in the weak $U(1)$ symmetry case (it is no longer gapped). Thus we return to the $(\psi_+,\psi_-)$ basis to find the right ``low energy'' degrees of freedom.

We explicitly write down the Keldysh action for the effective Liouvillian $\mathcal{L}^{\text{eff}}_{n = 1/2}$ :
\begin{align}
iS_2 & = \int dt\,\sum_i \psi_{i+}\partial_t \psi_{i+}^*+\psi_{i-}^*\partial_t \psi_{i-}\nonumber\\
& -iJ\sum_{\langle ij\rangle}(\psi_{i+}^{*}\psi_{j+}^{*}+\psi_{i+} \psi_{j+})+iJ\sum_{\langle ij\rangle}(\psi_{i-}^{*}\psi_{j-}^{*}+\psi_{i-} \psi_{j-})\nonumber\\
&+\Gamma\sum_{\langle ij\rangle}\bigg(2\psi_{i+}\psi_{j+}\psi_{i-}^*\psi_{j-}^*-|\psi_{i+}\psi_{j+}|^2-|\psi_{i-}\psi_{j-}|^2\bigg),
\end{align}
where $\psi_{i\pm}(t)$ are the bosonic fields at site $i$ on the forward (backward) branch of the closed time contour. Due to the transformation Eq.~(\ref{eq:largesn05}), the global $U(1)$ symmetry transformations that leave $S_2$ invariant are now:
\begin{align}
& \psi_{i+}\rightarrow \psi_{i+}e^{i\phi_+},\,i\in A;\quad \psi_{i+}\rightarrow \psi_{i+}e^{-i\phi_+},\,i\in B;\nonumber\\
& \psi_{i-}\rightarrow \psi_{i-}e^{i\phi_-},\,i\in A;\quad \psi_{i-}\rightarrow \psi_{i-}e^{-i\phi_-},\,i\in B;
\end{align}
At the saddle points $\delta S_2/\delta \psi_{i\pm} = 0$, the steady state fields $\psi_{i\pm}$ are uniform on each sublattice:
\begin{equation}
\psi_{A+}\psi_{B+} = -\frac{iJ}{\Gamma} = \psi_{A-}\psi_{B-},\quad |\psi_{A+}| = |\psi_{A-}|;
\label{eq:mf s2}
\end{equation}
where the latter constraint comes from higher order terms in the large spin expansion. Eq.~(\ref{eq:mf s2}) is in the symmetry broken phase, where both $U(1)$ symmetries ($U(1)$ rotation of $\psi_{-}$ and $\psi_+$) are  broken. As a result, the global phases of $\psi_+$ and $\psi_-$ are not determined in the mean field Eq.~(\ref{eq:mf s2}). Also, the relative amplitude of the fields $|\psi_A/\psi_B|$  on $A/B$ sublattice is not determined in Eq.~(\ref{eq:mf s2}). These degeneracies in mean field imply phase fluctuations of $\psi_{\pm}$  and the classical part of density fluctuation is gapless in $S_2$.

Consider fluctuations beyond mean field:
\begin{equation}
\psi_{i\pm} = \sqrt{\rho_{i\pm}+\delta\rho_{i\pm}}e^{i(\phi_{A/B\pm}+\delta\phi_{i\pm})},
\end{equation}
where we take $\rho_{i\in A\pm} = \rho_{i\in B\pm} = \sqrt{\frac{J}{\Gamma}}, \phi_{A-}+\phi_{B-} = \phi_{A+}+\phi_{B+} = -\frac{\pi}{2}$, and define the following new fields:
\begin{align}
& u_{i\pm} = (-1)^{s_i}\delta\rho_{i\pm}/\rho_{i\pm},\quad \delta\theta_{i\pm} = (-1)^{s_i}\delta\phi_{i\pm};\nonumber\\
&u_{i,c/q}= \frac{1}{\sqrt{2}}(u_{i+}\pm u_{i-}),\quad \theta_{i,c/q}= \frac{1}{\sqrt{2}}(\delta\theta_{i+}\pm \delta\theta_{i-});
\end{align}
where $s_i = 0(1)$ for $A(B)$ sublattice, the $u_{\pm},\theta_{\pm}$ fields represent density and phase fluctuations around the saddle point. We expand the action to quadratic order in $u,\theta$. And the action takes the form: $S_2 = S[u_q,\theta_c]+S[u_c,\theta_q]$. As is evident from the saddle point solution Eq.~(\ref{eq:mf s2}), fluctuations $\theta_c,\theta_q,u_c$ are gapless, and $u_q$ is gapped with a mass term. We integrate out $u_q$ and arrive at an effective action $S_2^{\text{eff}}$ :
\begin{align}
&S_2^{\text{eff}} = S[\theta_c]+S[u_c,\theta_q],\nonumber\\
&S[\theta_c] = \frac{i}{2\gamma}\int dt d^dx\,(\partial_t\theta_c-D\nabla^2\theta_c)^2,\nonumber\\
&S[u_c,\theta_q] = \int dt d^dx\,(-u_c\partial _t\theta_q+D\nabla u_c\nabla\theta_q)+i\sigma(\nabla\theta_q)^2,
\label{eq:eff action s}
\end{align}
where the phenomenological coupling constants are determined by parameters in the microscopic model. Detailed derivation is provided in Appendix \ref{app:derivation}. As expected, the effective action $S_2^{\text{eff}}$ respects two $U(1)$ symmetries ($U(1)$ rotation of $e^{i\theta_c}$ and $e^{i\theta_q}$) and they are encoded in two parts: $S[\theta_c]$ describes the spontaneous breaking of weak $U(1)$ symmetry, and $S[u_c,\theta_q]$ describes the $U(1)$ strong-to-weak symmetry breaking. 

$S[\theta_c]$ describes the fluctuation of weak $U(1)$ symmetry order parameter $e^{i\theta_c}$. It is of the same form as the effective action for $S_1$ in Sec.~\ref{subsec:keldysh w}, which corresponds to a Langevin equation for $\theta_c$:
\begin{equation}
\partial_t\theta_c = D\nabla^2\theta_c + \eta(x,t),
\label{eq:diff}
\end{equation}
where long-range order is preserved in $d\geq 3$. In low dimensions $d\leq 2$, long-range order is destroyed by fluctuations, and similar to Eq.~(\ref{eq:kpz}), a nonlinear term $\sim (\nabla\theta_c)^2$ beyond our quadratic approximation becomes relevant, and there is even no quasi long-range order.

For $S[u_c,\theta_q]$, we can integrate by part to arrive at an action describing the diffusive hydrodynamics of density fluctuation $u_c$:
\begin{equation}
S[u_c,\theta_q] = \int dt d^dx\,i\sigma(\nabla\theta_q)^2+\theta_q(\partial_t u_c-D\nabla^2 u_c),
\end{equation}
from which we can calculate the correlation:
\begin{equation}
\langle e^{i\theta_q(x,t)}e^{-i\theta_q(0,0)}\rangle \sim \text{Const},\  \forall d.
\label{eq:lro eff}
\end{equation}
The long-range correlation of $e^{i\theta_q}$ dictates the strong-to-weak symmetry breaking, which occurs in any dimension. It is equivalent to the R\'enyi-2 correlator previously studied in \cite{lee2023quantum,lessa2024strong,sala2024spontaneous}, both describing the long-range correlation of the strong-to-weak SSB order parameter. The effective action we derived is quite general and has been used as the lowest-order phenomenological description of diffusive hydrodynamics in interacting classical and quantum systems \cite{michailidis2023corrections,Effectivefieldtheory}. Here we derive the action from a concrete microscopic model Eq.~(\ref{mod:strong}) and interpret it as describing $U(1)$ strong-to-weak SSB. In fact, the density-density correlation obeys the following form:
\begin{equation}
\langle u_c(x,t)u_c(0,0)\rangle \sim \frac{1}{\sqrt{4\pi Dt}}e^{-\frac{x^2}{4Dt}}.
\end{equation}
We indeed see that strong-to-weak $U(1)$ symmetry breaking leads to gapless diffusion modes of the symmetry charge. 

For $S_2$, there are two spontaneous symmetry-breaking processes: not only the strong $U(1)$ symmetry breaks to the corresponding weak $U(1)$, but the weak $U(1)$ symmetry also spontaneously breaks. Consequently, there are two Goldstone modes, corresponding to spin waves and diffusion of symmetry charge, respectively.

\section{\label{sec:s-to-w} ingappability from Strong-to-Weak symmetry breaking}
In this section, we elaborate more on how the interplay between strong $U(1)$ symmetry and translational invariance leads to gapless Liouvillian spectrum at any filling. In Sec.~\ref{subsec:keldysh s}, we have seen that in the effective action $S_2^{\text{eff}}$ of $\mathcal{L}_{\text{II}}$ near half filling, there is a part $S[u_c,\theta_q]$ associated with strong-to-weak symmetry breaking that leads to gapless diffusion modes of density fluctuation. Here through a simple model, we establish a connection between gapless Goldstone modes of strong-to-weak symmetry breaking and diffusion of the corresponding symmetry charge at general filling factors (except for the empty or the full filling case).

%There are a few remarks we want to make before our discussion. First, LSM theorems in closed quantum systems \cite{LIEB1961407, PhysRevLett.84.1535} imply that with $U(1)$ symmetry and translational invariance, the ground states at fractional filling are nontrivial. For Liouvillians with strong $U(1)$ symmetry and translational invariance, the spectrum is always gapless, but the steady states can be trivial, for example, if the dissipators are hermite, the steady states are simply identities in each symmetry sector. 

For systems at ground states or in thermal equilibrium, spontaneous symmetry breaking takes place only for certain parameter regime (or below critical temperature $T_c$). Moreover, in low dimensions, it is usually the case that fluctuations are so strong as to destroy SSB. However, strong-to-weak symmetry breaking always takes place for any Liouvillian with strong symmetry in any dimensions. This is 
guaranteed by the structure of strongly symmetric Liouvillians: One can always find steady states that breaks the strong symmetry for strongly symmetric Liouvillians (as discussed in Sec.~\ref{sec:sym}).

The model we consider here is a spin-S XXZ chain under dephasing:
\begin{align}
& H_{\text{XXZ}} = \sum_i\bigg[\frac{J_{xy}}{2}(S_i^+S_{i+1}^-+S_i^-S_{i+1}^+)+J_zS_i^zS_{i+1}^z\bigg],\nonumber\\
& \mathcal{L}_{\text{III}
}[\rho]= -i[H_{\text{XXZ}},\rho] + \Gamma\sum_i\mathcal{D}(S_i^z)[\rho],
\label{eq:xxz deph}
\end{align}
where $S_i^{\pm,z}$ are the spin operators. The $S = 1/2$ case has been studied in \cite{PhysRevLett.111.150403}. 

In \cite{kawabata2023lieb}, it is argued from a LSM point of view that the $S = 1/2$ model is gapless while a dissipative gap can be open for $S = 1$. However, here we demonstrate, for general spin and at general filling, model III is always gapless from strong-to-weak symmetry breaking and the Goldstone modes in the physical space correspond to diffusion of the symmetry charge.

$\mathcal{L}_{\text{III}}$ respects the strong $U(1)$ symmetry generated by $S^z = \sum_i S_i^z$ or $N = \sum_i n_i = \sum_i (S_i^z+S)$ if we treat the spins as bosons. The steady states of $\mathcal{L}_{\text{III}}$ are trivial, which are maximally mixed states $I^{S^z}$ in each $S^z$ sector. Thus, it is clear that the weak U(1) symmetry is not spontaneously broken:
\begin{equation}
\mathcal{U}^w(\theta)[I^{S^z}] = e^{-i\hat{S}^z\theta}I^{S^z}e^{i\hat{S}^z\theta} = I^{S^z}.
\end{equation}
However, the strong $U(1)$ symmetry spontaneously breaks to weak $U(1)$ symmetry since a superposition of $I^{S_z}$'s is still a steady state, but it is not invariant under strong $U(1)$ symmetry operation Eq.~(\ref{eq:stro}):
\begin{equation}
\mathcal{U}^s(\theta)[\sum_{S^z} I^{S^z}] = e^{-i\hat{S}^z\theta}\sum_{S^z} I^{S^z} = \sum_{S^z} e^{-iS^z\theta}I^{S^z}.
\end{equation}
It is also straightforward to see that $I^{S_z}$ are long-range ordered in the sense that the R\'enyi-2 correlator Eq.~(\ref{lro:swssb}) calculated from $I^{S^z}$ is a non-zero constant independent of $|i-j|$. We show below that strong-to-weak symmetry breaking renders the spectrum of $\mathcal{L}_{\text{III}}$ gapless with Goldstone modes corresponding to diffusion of the conserved symmetry charge $\langle S_i^z\rangle$.

When there is no Hamiltonian term $J_{xy} = J_z = 0$, the steady states are classical superpositions of the symmetry charge $S_i^z$ eigenstates:
\begin{align}
&\rho_{ss} = \sum_{\{S_i^z\}}p_{\{S_i^z\}}|\{S_i^z\}\rangle\langle\{S_i^z\}|,\nonumber\\
&\{S_i^z\} = \{S_1^z,S_2^z\cdots\},\quad S_i^z = -S,-S+1\cdots S;
\label{eq:subspace}
\end{align}
where $p_{\{S_i^z\}}\geq 0 $  and $\sum_{\{S_i^z\}}p_{\{S_i^z\}} = 1$.

Now we consider adding $H_{\text{XXZ}}$ perturbatively. For the effective Liouvillian $\mathcal{L}_{\text{III}}^{\text{eff}}$, we consider only density matrices in the original steady state subspace. By adopting the following map
\begin{equation}
|\{S_i^z\}\rangle\langle\{S_i^z\}|\rightarrow |\{S_i^z\}\rangle,
\end{equation}
 $-\mathcal{L}_{\text{III}}^{\text{eff}}$ takes the form of a spin chain Hamiltonian $H^{\text{eff}}$, with ground state energy being $0$. ($\mathcal{L}_{\text{III}}^{\text{eff}}$ is Hermitian because the dissipators are Hermitian, and there is no odd-order contribution of $H_{\text{XXZ}}$). Specifically, in the case of $S = 1/2$, to second order in $J_{xy}$, $H^{\text{eff}}$ is the ferromagnetic Heisenberg model Hamiltonian \cite{PhysRevLett.111.150403}:
\begin{align}
&H^{\text{eff}} = \frac{J_{xy}^2}{2\Gamma}\sum_i\bigg[\frac{1}{2}(S_i^+S_{i+1}^-+h.c.)+S_i^zS_{i+1}^z-\frac{1}{4}\bigg].
\label{eq:eff 3}
\end{align}
 As discussed in Sec.~\ref{sec:sym}, the strong $U(1)$ symmetry has two generators in the doubled space: $N_L-N_R$ and $\frac{1}{2}(N_L+N_R)$. For the effective Liouvillian $\mathcal{L}_{\text{III}}^{\text{eff}}$ (or $H^{\text{eff}}$), $N_L-N_R$ generated $U(1)$ symmetry acts trivially as identity and $\frac{1}{2}(N_L+N_R)$ generated $U(1)$ symmetry is the $U(1)$ rotation around the $z$ axis. Moreover, the $U(1)$ symmetry of $H^{\text{eff}}$ is spontaneously broken. The ground states in different $S^z$ sectors are all degenerate with $E_G = 0$: $H^{\text{eff}}|I^{S^z}\rangle = 0$, where $|I^{S^z}\rangle$ is the doubled space map of $I^{S^z}$:
 \begin{align}
|I^{S^z}\rangle \sim\sum_{\sum_iS_i^z = S^z}|\{S_i^z\}\rangle,
\end{align}
which corresponds to magnon condensation at $k = 0$. 

 The above discussion on spontaneous $U(1)$ symmetry breaking of $H^{\text{eff}}$ holds true for general $S$ and to arbitrary order in the perturbation theory. $H^{\text{eff}}$ is always in the $U(1)$ symmetry broken phase, with magnon condensation in the ground states. To illustrate the corresponding gapless Goldstone modes in each symmetry sector, we write down variational wave functions in each $S^z$ sector orthogonal to $|I^{S^z}\rangle$, with vanishing energy in the thermodynamical limit. By mapping the spins to bosons:
 \begin{align}
&n_i = b^{\dagger}_ib_i = S_i^z+S,\quad b_i^{\dagger}|n-S\rangle_i= \sqrt{n+1}|n+1-S\rangle_i,\nonumber\\
&n_i\leq 2S,\quad b_i^{\dagger}|S\rangle_i= 0;
\end{align}
the steady states (or ground states of $H^{\text{eff}}$) satisfy:
\begin{align}
& I^{S^z} = \frac{1}{N}\sum_i b_i^{\dagger}I^{S^z-1}b_i,\nonumber\\
& |I^{S^z}\rangle = \frac{1}{N}\sum_i b_i^{\dagger}|I^{S^z-1}\rangle;
\end{align}
 where $N = \sum n_i = \sum S_i^z + S$. Then the variational wave function in the form of Goldstone modes are:
 \begin{align}
& |k\rangle \sim\sum_r e^{ikr}\,b_r^{\dagger}|I^{S^z-1}\rangle,\nonumber\\
&\rho_k \sim\sum_r e^{ikr}\,b_r^{\dagger}I^{S^z-1}b_r.
\end{align}
Since in the ground states $|I^{S^z}\rangle$, the magnons condense at $k = 0$, the variational wave function $|k\rangle$ gives one of the magnons a nonzero momentum $k$. One can check that $|k\rangle$ is orthogonal to $|I^{S^z}\rangle$ and in the limit $k\rightarrow 0$, $|k\rangle$ has vanishing energy $\langle k|H^{\text{eff}}|k\rangle\rightarrow0$.

In the physical Hilbert space, the effective Liouvillian $\mathcal{L}_{\text{III}}^{\text{eff}}$ governs the evolution of the density matrix in the subspace defined by Eq.~(\ref{eq:subspace}), which is nothing but the evolution of the probability distribution $p_{\{S_i^z\}}$. The Goldstone modes in the physical Hilbert space can pair in opposite momentum:
\begin{align}
&\rho^k + \rho^{-k}  \sim \sum_r \cos kr\,b_r^{\dagger}I^{S^z-1}b_r,\nonumber\\
&\rho^k - \rho^{-k}  \sim \sum_r \sin kr\,b_r^{\dagger}I^{S^z-1}b_r;
\end{align}
 which corresponds to the density fluctuation of the symmetry charge $\langle S_i^z\rangle$(or $\langle n_i^z\rangle$). The dispersion is proportional to $k^2$ (this is evident from the one particle sector):
 \begin{equation}
\lambda_k\sim k^2,
 \end{equation}
where $\lambda_k$ is the Liouvillian gap in each symmetry sector. Consequently, the Goldstone modes of strong-to-weak symmetry breaking govern the diffusion of the conserved symmetry charge. The connection between gapless goldstone modes and diffusion has also been studied in the context of Brownian random time evolution \cite{PhysRevLett.131.220403} and symmetry algebras in \cite{moudgalya2023symmetries}.

Now that we have constructed gapless diffusion modes at any filling factor and for general spin, it is clear how strong-to-weak symmetry breaking leads to the gapless spectrum of $\mathcal{L}_{\text{III}}$. Generally, we expect that the spectrum of a Liouvillian with strong $U(1)$ symmetry and translational invariance at general filling is always gapless due to strong-to-weak SSB.

\section{\label{sec:conc} Concluding remarks}
We have studied symmetries and their spontaneous breaking in open quantum systems. Unlike closed systems, open systems display two distinct types of symmetries: strong and weak, depending on whether there is an exchange of symmetry charges with the environment. We outline different scenarios of spontaneous symmetry breaking in open systems and provide concrete examples for each scenario. For strong continuous symmetry, there are two symmetry-breaking processes: the inevitable strong-to-weak symmetry breaking leads to diffusion of symmetry charge, and possible further weak symmetry breaking leads to long-range order with Goldstone modes describing fluctuations of the order parameter. We also uncover a novel transition,  tuned by filling, from a symmetric phase with a Bose surface to a symmetry-broken phase with long-range order.

There are a number of interesting future directions to explore. First, whereas we have shown how strong-to-weak symmetry breaking leads to a gapless Liouvillian spectrum and diffusion modes of the conserved charge in several concrete models, there are scenarios where the Liouvillian gap does not faithfully reflect the relaxation time of observables \cite{Song19chiral,Mori20resolve,Haga21Liouvillian,Lee23anomalously,rakovszky2023defining}. For a specific fine-tuned Liouvillian with strong continuous symmetry and translational invariance, at a certain symmetry charge sector, the Liouvillian gap may stay open while relaxation time diverges. It is desirable to find a quantity that faithfully reflects the relaxation time of physical observables. Second, as for symmetries in open quantum systems, we consider here only the simplest $U(1)$ case. It would be interesting to extend the results to other symmetries, for example, non-abelian symmetries and their constraints on the dynamics and steady states \cite{lessa2024symmetry,pollmann2024highly}.

\emph{Note added.}--- When we were updating the second version of the manuscript, a preprent appeared with partial overlap with our work \cite{huang2024hydro}. In their work, an effective field theory of strong-to-weak symmetry breaking is also studied.

\begin{acknowledgments}
{\it Acknowledgments}.---We thank Zi Cai for helpful discussions. This work is supported by the NSFC under Grant No. 12125405,  National Key R\&D Program of China (No. 2023YFA1406702),  and the Innovation Program for Quantum Science and Technology (Grant No. 2021ZD0302502).
%and the Innovation Program for Quantum Science and Technology (No. 2021ZD0302502). 
\end{acknowledgments}

\appendix
\begin{widetext}

\section{ Perturbative analysis of Model I}\label{app:perturbation}
In this appendix we provide the derivation the effective Liouvillian $\mathcal{L}_I^{\text{eff}}$ Eq.~(\ref{equ:eff liou I}). For $\mathcal{L}_{I}$ Eq.~(\ref{mod:weak}), we still take $J_z = 0$ and treat the Hamiltonian part as perturbation. For the dissipation part:
\begin{equation}
\mathcal{L}^0[\rho] = \Gamma\sum_{i\in A}\mathcal{D}(\sigma_i^-)[\rho]+\Gamma\sum_{i\in B}\mathcal{D}(\sigma_i^+)[\rho], 
\end{equation} 
all the eigenvalues and eigenstates can be easily obtained since the dissipators are on site. Consider a single site problem with only loss: $\mathcal{L}[\rho] = \Gamma\mathcal{D}(\sigma^-)[\rho]$, the eigenvalues and eigenstates are shown in Table.~(\ref{tab:eigen}):
\begin{table}[h]
\begin{tabular}{ |p{3cm}|p{3cm}|p{3cm}|  }
 \hline
 Eigenvalue & Left Eigenstate $e^L$ & Right Eigenstate $e^R$\\
 \hline
0           & $I = |0\rangle\langle0 |+|1\rangle\langle 1|$ & $|0\rangle\langle0|$\\
 \hline
$-\Gamma$   & $|1\rangle\langle 0|$                   & $|1\rangle\langle 0|$\\
 \hline
$-\Gamma$   & $|0\rangle\langle 1|$                   & $|1\rangle\langle 0|$\\
 \hline 
$-2\Gamma$  & $|1\rangle\langle 1|$                   & $|1\rangle\langle 1|-|0\rangle\langle 0|$\\
 \hline
\end{tabular}
\caption{Eigenvalues and eigenstates of a a single site problem with only loss. Here $0$ stands for spin down and $1$ stands for spin up. The eigenstates are biorthonormal: $\tr(e_m^{L\dagger}e_n^R) = \delta_{mn}$.} 
\label{tab:eigen}
\end{table} 

For a single site problem with only gain, the eigenvalues and eigenstates are the same except that $0$ and $1$ are exchanged. For the unperturbed Liouvillian $\mathcal{L}^0$, the eigenstates are simply direct product of eigenstates of each site. For example, the steady state of $\mathcal{L}^0$ is $\rho_0 = \bigotimes_{i\in A}|0\rangle\langle 0|_i\bigotimes_{j\in B}|1\rangle\langle 1|_j$. The corresponding left eigenstate is $\bigotimes_i |0\rangle\langle0 |_i+|1\rangle\langle 1|_i$.

When $H$ is added perturbatively, we focus on the space in symmetry sector $N_L-N_R = 1$ with eigenvalue $-\Gamma$. This space is spanned by the (right) eigenstates of $\mathcal{L}^0$ :
\begin{equation}
e^R_i = \left \{
\begin{array}{ll}
    \sigma_i^+\rho_0,     & i\in A\\ 
    \rho_0\sigma_i^+,     & i\in B
\end{array}
\right.
\end{equation}
we use $i$ to label them. The corresponding left eigenstates $e_i^L$ can be obtained as discussed above. To first order in perturbation theory, we only need to know how $\delta \mathcal{L}[\cdot] = -i[H,\cdot]$ acts in this space, which amounts to calculate the matrix elements $\delta \mathcal{L}_{ij} = \tr (e_i^{L\dagger}\delta \mathcal{L}[e_j^R])$. $\delta\mathcal{L}_{ij}$ has a tight binding form, it is nonzero only when $i,j$ are nearest neighbour:
\begin{equation}
\delta\mathcal{L}_{\langle ij\rangle} = 
\left \{
\begin{array}{ll}
    -iJ,     & i\in A,\,j\in B\\ 
     iJ.     & i\in B,\,j\in A
\end{array}
\right.
\end{equation}
We illustrate this with a simple one dimensional example in Fig.~(\ref{fig:perturbation}):
\begin{figure}[h]
    \centering
    \includegraphics[width=17cm]{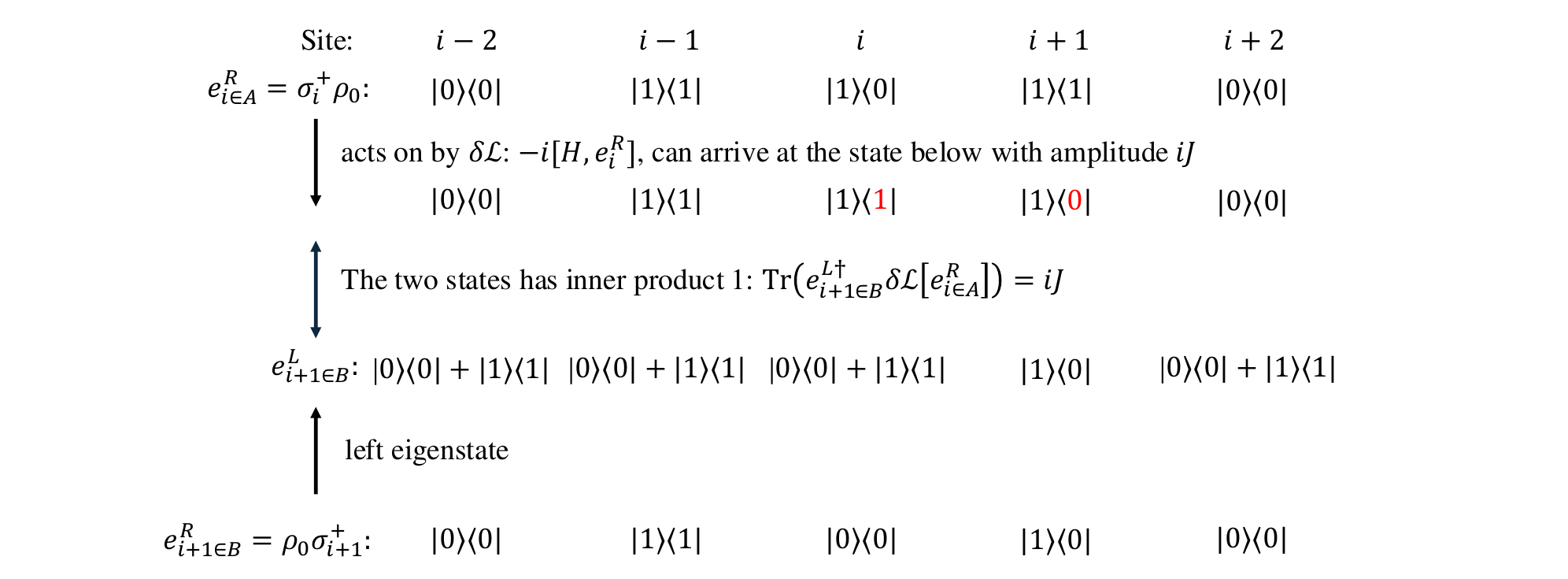}
    \caption{Calculation of the matrix elements $\delta\mathcal{L}_{ij}$. The figure illustrates the case where $\delta\mathcal{L}_{\langle BA\rangle} = iJ$; $\delta\mathcal{L}_{\langle AB\rangle} = -iJ$ can be calculated similarly.}
    \label{fig:perturbation}
\end{figure}

Finally, the effective Liouvillian in this subspace has the matrix form:
\begin{align}
\mathcal{L}^{\text{eff}}_{\text{I}} & = \mathcal{L}^0 + \delta\mathcal{L}\nonumber\\
& = -\Gamma + iJ\sum_{\langle ij\rangle} |i\rangle\langle j|-|j\rangle\langle i|,\quad(i\in B, j\in A).
\end{align}
\section{ Derivation of the effective action for strong symmetric Model II}\label{app:derivation}

In this appendix, we derive the effective action Eq.~(\ref{eq:eff action s}) for the strong $U(1)$ symmetric model $\mathcal{L}_{\text{II}}$ Eq.~(\ref{mod:strong}) near $n = 1/2$, starting from the bosonic Liouvillian $\mathcal{L}^{\text{eff}}_{n = 1/2}$ Eq.~(\ref{eq:eff liou 05}). The Keldysh action for $\mathcal{L}^{\text{eff}}_{n = 1/2}$:
\begin{align}
iS_2 & = \int dt\,\sum_i \psi_{i+}\partial_t \psi_{i+}^*+\psi_{i-}^*\partial_t \psi_{i-} -iJ\sum_{\langle ij\rangle}(\psi_{i+}^{*}\psi_{j+}^{*}+\psi_{i+} \psi_{j+})+iJ\sum_{\langle ij\rangle}(\psi_{i-}^{*}\psi_{j-}^{*}+\psi_{i-} \psi_{j-})\nonumber\\
&+\Gamma\sum_{\langle ij\rangle}\bigg(2\psi_{i+}\psi_{j+}\psi_{i-}^*\psi_{j-}^*-|\psi_{i+}\psi_{j+}|^2-|\psi_{i-}\psi_{j-}|^2\bigg),
\end{align}
where there is a $U(1)\times U(1)$ symmetry, which are $U(1)$ rotations of $\psi_{\pm}$:
\begin{align}
& \psi_{i+}\rightarrow \psi_{i+}e^{i\phi_+},\,i\in A;\quad \psi_{i+}\rightarrow \psi_{i+}e^{-i\phi_+},\,i\in B;\nonumber\\
& \psi_{i-}\rightarrow \psi_{i-}e^{i\phi_-},\,i\in A;\quad \psi_{i-}\rightarrow \psi_{i-}e^{-i\phi_-},\,i\in B;
\end{align}
We separate the imaginary part and the real part of the action:
\begin{align}
&iS_2 = \int dt\,(\text{im})\bigg[\sum_i \psi_{i+}\partial_t \psi_{i+}^*+\psi_{i-}^*\partial_t \psi_{i-} -iJ\sum_{\langle ij\rangle}(\psi_{i+}^{*}\psi_{j+}^{*}+\psi_{i+} \psi_{j+})+iJ\sum_{\langle ij\rangle}(\psi_{i-}^{*}\psi_{j-}^{*}+\psi_{i-} \psi_{j-})\nonumber\\
&+\Gamma\sum_{\langle ij\rangle}(\psi_{i+}\psi_{j+}\psi_{i-}^*\psi_{j-}^*-\psi_{i+}^*\psi_{j+}^*\psi_{i-}\psi_{j-})\bigg]-(\text{re})\bigg[\Gamma\sum_{\langle ij\rangle}|a_{m+}a_{n+}-a_{m-}a_{n-}|^2\bigg],
\end{align}
the Euler-Lagrange equation (saddle point for the imaginary part of $iS$) is:
\begin{align}
&\partial_t \psi_{i+} = \sum_{\text{nearest j}}-iJ \psi_{j+}^*-\Gamma \psi_{j+}^*\psi_{j-}\psi_{i-},\nonumber\\
&\partial_t \psi_{i-} = \sum_{\text{nearest j}}-iJ \psi_{j-}^*-\Gamma \psi_{j-}^*\psi_{j+}\psi_{i+};
\end{align}
the steady state satisfies:
\begin{equation}
\psi_{A+}\psi_{B+} = -\frac{iJ}{\Gamma} = \psi_{A-}\psi_{B-},
\label{eq:steady state}
\end{equation}
where the fields on $A/B$ sublattice are uniform and represented by $\psi_{A/B\pm}$. In the steady state, the global phases of $\psi_{\pm}$ (relative phase between A/B sublattice) are free: the global phase can take any value, which is the consequence of $U(1)\times U(1)$ symmetry of the action. The relative density between $A/B$ sublattice: $|\psi_A|^2-|\psi_B|^2$ is also free, for both $\psi_{\pm}$. However, as calculation turns out, if we consider higher order terms in the large spin expansion, in the steady state it is also required:
\begin{equation}
|\psi_{A+}| = |\psi_{A-}|,\quad |\psi_{B+}| = |\psi_{B-}|;
\end{equation}
which means that only the classical part of the density fluctuation $\rho_{Ac}-\rho_{Bc} = (|\psi_{A+}|^2+|\psi_{A-}|^2)-(|\psi_{B+}|^2+|\psi_{B-}|^2)$ is now free in the steady state (this is due to the conservation of $\sum_{i\in A}b_i^{\dagger}b_i-\sum_{i\in B}b_i^{\dagger}b_i$), the quantum part of density $\rho_q\sim\rho_{+}-\rho_{-}$ is simply 0 in the steady state.

Written in terms of density and phase $\psi_{i\pm} = \sqrt{\rho_{i\pm}}e^{i\theta_{\pm}}$, the action is:
\begin{align}
iS_2 & = \int dt\,i\sum_i -\rho_{i+}\partial_t\theta_{i+}+\rho_{i-}\partial_t\theta_{i-} -2iJ\sum_{\langle ij\rangle}\sqrt{\rho_{i+}\rho_{j+}}\cos(\theta_{i+}+\theta_{j+})+2iJ\sum_{\langle ij\rangle}\sqrt{\rho_{i-}\rho_{j-}}\cos(\theta_{i-}+\theta_{j-})\nonumber\\
&+2i\Gamma\sum_{\langle ij\rangle}\sqrt{\rho_{i+}\rho_{j+}\rho_{i-}\rho_{j-}}\sin(\theta_{i+}+\theta_{j+}-\theta_{i-}-\theta_{j-})-\Gamma\sum_{\langle ij\rangle}|\sqrt{\rho_{i+}\rho_{j+}}e^{i(\theta_{i+}+\theta_{j+})}-\sqrt{\rho_{i-}\rho_{j-}}e^{i(\theta_{i-}+\theta_{j-})}|^2,
\end{align}
Consider fluctuations around saddle point $\psi_{i\pm} = \sqrt{\rho_{i\pm}+\delta\rho_{i\pm}}\,e^{i(\theta_{i\pm}+\delta\theta_{i\pm})}$, where we choose $\theta_{A-}+\theta_{B-} = \theta_{A+}+\theta_{B+} = -\pi/2$ and $\rho_{A-}\rho_{B-} = \rho_{A+}\rho_{B+} = J^2/\Gamma^2$. To quadratic order in $\delta\rho$ and $\delta\theta$: 
\begin{align}
iS_2 & = \int dt\,i\sum_i -\delta\rho_{i+}\partial_t\delta\theta_{i+}+\delta\rho_{i-}\partial_t\delta\theta_{i-}\nonumber\\
& -2i\frac{J^2}{\Gamma}\sum_{\langle ij\rangle}(\frac{\delta\rho_{i+}}{2\rho_{i+}}+\frac{\delta\rho_{j+}}{2\rho_{j+}})(\delta\theta_{i-}+\delta\theta_{j-})+2i\frac{J^2}{\Gamma}\sum_{\langle ij\rangle}(\frac{\delta\rho_{i-}}{2\rho_{i-}}+\frac{\delta\rho_{j-}}{2\rho_{j-}})(\delta\theta_{i+}+\delta\theta_{j+})\nonumber\\
&-\frac{J^2}{\Gamma}\sum_{\langle ij\rangle}(\frac{\delta\rho_{i+}}{2\rho_{i+}}+\frac{\delta\rho_{j+}}{2\rho_{j+}}-\frac{\delta\rho_{i-}}{2\rho_{i-}}-\frac{\delta\rho_{j-}}{2\rho_{j-}})^2+(\delta\theta_{i+}+\delta\theta_{j+}-\delta\theta_{i-}-\delta\theta_{j-})^2,
\end{align}
as discussed above, when higher order terms in large spin expansion are considered, at saddle point we have $\rho_{i+} = \rho_{i-}$. Here we consider the simplest scenario $\rho_{A+} = \rho_{A-} = \rho_{B+} = \rho_{B-} = J/\Gamma$ (the case where $\rho_A\neq \rho_B$ are similar qualitatively). And we also make variable change: $u_i = (-1)^i\delta\rho_i/\rho_i$, $\delta\theta_i = (-1)^i\delta\theta_i$ ($i = 0/1$ for A/B sublattice). Then the action takes a simpler form:
 \begin{align}
iS_2 & = \int dt\,i\frac{J}{\Gamma}\sum_i - u_{i+}\partial_t\delta\theta_{i+}+ u_{i-}\partial_t\delta\theta_{i-}\nonumber\\
& -i\frac{J^2}{\Gamma}\sum_{\langle ij\rangle}( u_{i+}- u_{j+})(\delta\theta_{i-}-\delta\theta_{j-})+i\frac{J^2}{\Gamma}\sum_{\langle ij\rangle}( u_{i-}- u_{j-})(\delta\theta_{i+}-\delta\theta_{j+})\nonumber\\
&-\frac{J^2}{\Gamma}\sum_{\langle ij\rangle}\frac{1}{4}(u_{i+}-u_{j+}-u_{i-}+u_{j-})^2+(\delta\theta_{i+}-\delta\theta_{j+}-\delta\theta_{i-}+\delta\theta_{j-})^2,
\end{align}
in terms of classical and quantum variable $u_{ic/q} = \frac{1}{\sqrt{2}}(u_{i+}\pm u_{i-}),\theta_{ic/q} = \frac{1}{\sqrt{2}}(\delta\theta_{i+}\pm \delta\theta_{i-})$:
\begin{align}
iS_2 & = \int dt\,-i\frac{J}{\Gamma}\sum_i  u_{ic}\partial_t\theta_{iq}+u_{iq}\partial_t\theta_{ic}\nonumber\\
&+i\frac{J^2}{\Gamma}\sum_{\langle ij\rangle}( u_{ic}- u_{jc})(\theta_{iq}-\theta_{jq})-i\frac{J^2}{\Gamma}\sum_{\langle ij\rangle}( u_{iq}- u_{jq})(\theta_{ic}-\theta_{jc})\nonumber\\
&-\frac{J^2}{\Gamma}\sum_{\langle ij\rangle}\frac{1}{2}(u_{iq}-u_{jq})^2+(\theta_{iq}-\theta_{jq})^2,
\end{align}
in the above action, there is no mass term, which means all the fluctuations $u_{c/q},\theta_{c/q}$ are gapless. However, as discussed above, with higher order terms, the quantum part of density fluctuation $u_q$ should be gapped. The correct action takes the form:
\begin{align}
iS_2 & = \int dt\,-i\frac{J}{\Gamma}\sum_i  u_{ic}\partial_t\theta_{iq}+u_{iq}\partial_t\theta_{ic}\nonumber\\
&+i\frac{J^2}{\Gamma}\sum_{\langle ij\rangle}( u_{ic}- u_{jc})(\theta_{iq}-\theta_{jq})-i\frac{J^2}{\Gamma}\sum_{\langle ij\rangle}( u_{iq}- u_{jq})(\theta_{ic}-\theta_{jc})\nonumber\\
&-\frac{J^2}{\Gamma}\sum_{\langle ij\rangle}\frac{1}{2}(u_{iq}-u_{jq})^2+(\theta_{iq}-\theta_{jq})^2-\frac{m_q^2}{2}\sum_iu_{iq}^2,
\end{align}
where $m_q$ arises from higher order terms in large spin expansion. Then it is straightforward to go to the continuum limit and integrate out the gapped $u_q$ fluctuation:
\begin{align}
&S_2^{\text{eff}} = S[\theta_c]+S[u_c,\theta_q],\nonumber\\
&S[\theta_c] = \frac{i}{2\gamma}\int dt d^dx\,(\partial_t\theta_c-D\nabla^2\theta_c)^2,\nonumber\\
&S[u_c,\theta_q] = \int dt d^dx\,(-u_c\partial _t\theta_q+D\nabla u_c\nabla\theta_q)+i\sigma(\nabla\theta_q)^2,
\end{align}
where the parameters $\gamma,D,\sigma$ arises from microscopic parameter $J,\Gamma,m_q$.
\end{widetext}

% Create the reference section using BibTeX
\bibliography{references}

\end{document}